\documentclass[11 pt]{article}
\usepackage[utf8]{inputenc}
\usepackage{amsmath,enumerate}

\newtheorem{asml}{Assumption S}

\usepackage{bm}
\usepackage{bbm}
\usepackage[margin=1in]{geometry}
\usepackage{natbib}
\usepackage{rotating}
\usepackage[usenames]{color}
\usepackage{xcolor}
\usepackage{comment}
\usepackage{url}
\usepackage{graphicx}
\usepackage{caption}
\usepackage{subcaption}
\usepackage{setspace}
\usepackage{rotating}
\usepackage{amssymb,amsmath,amsfonts}
\usepackage{mdsymbol,soul}
\usepackage{tikz}
\usetikzlibrary{bayesnet}

\graphicspath{ {} }

\begin{document}

\title{Latent Dirichlet Analysis of Categorical  Survey Responses}
\author{ Evan Munro \thanks{{\small Graduate School of Business, Stanford University. Email: munro@stanford.edu,}}\and Serena Ng \thanks{{\small Department of Economics, Columbia University, and NBER. Email: serena.ng@columbia.edu}\newline We thank Stephen Hansen, Guido Imbens,  Pepe Montiel Olea,  Rui Yu,   as well as the Gianni Amisano (discussant) and  participants at the {\em Nontraditional Data, Machine Learning, and Natural Language Processing in Macroeconomics} conference at the Federal Research Board  for many helpful comments. Financial support from the National Science Foundation (SES 1558623) is gratefully acknowledged. Code and data to replicate analysis in the paper is at \texttt{https://github.com/evanmunro/lda-survey-exp}.}}
\date{\today}

\maketitle
\begin{abstract}
Beliefs are important determinants of an individual's choices and economic outcomes, so understanding how they  comove and differ across individuals is of considerable interest. Researchers often rely on surveys that report individual beliefs as qualitative data. We propose using a Bayesian hierarchical latent class model to analyze the comovements and observed heterogeneity in categorical survey responses. We show that the statistical model corresponds to an economic structural model of information acquisition, which guides interpretation and estimation of the model parameters.  An algorithm based on stochastic optimization is proposed to estimate a model for repeated surveys when responses follow a dynamic structure and conjugate priors are not appropriate. Guidance on selecting the number of belief types is also provided. Two examples are considered. The first  shows that there is information in the Michigan survey responses beyond the consumer sentiment index that is officially published.  The  second shows that belief types constructed from survey responses can be used in a subsequent analysis to estimate heterogeneous returns to education. 

\end{abstract}



\thispagestyle{empty}
\setcounter{page}{0}
\newpage
\bibliographystyle{econometrica}
\section{Introduction}
An important part of economic analysis is understanding why  economic outcomes differ across individuals. Heterogeneity in individual beliefs is presumed to be an  important factor contributing to  the observed differences. But, as beliefs are not easily observed directly,  researchers often turn to  surveys that  solicit such  information and report the qualitative responses in  categorical form.  For example, the Michigan Consumer Survey asks if individuals  expect home prices to increase or decrease, and whether they believe business conditions are worse or better than the previous year. The Gallup Poll solicits sentiments towards politics, values and the environment, while the Bank of England Inflation Attitudes Survey tracks inflation expectations as categorical responses. Individuals' views on stock valuations have  been studied in \citet{shiller1996did}.  In this paper, we use the term `beliefs' to represent  qualitative assessments about unknown or unrealized quantities. This encompasses such economic terms  as  `expectations', which are  beliefs about future values of specific variables such as inflation rates,  `sentiments', which are  beliefs about current and future economic conditions in broad terms,  as well as `attitudes' and  `self-assessments', which are beliefs about  health, wealth, and other socio-economic dimensions. 

A distinctive feature of beliefs about the economy is that they are cross-sectionally correlated, presumably due to public information about the  economy, but the correlation is not perfect, since individuals are heterogeneous. The goal is to design a methodology that is grounded in economic theory and can analyze heterogeneous and qualitative responses. Existing econometric methods designed for analyzing categorical data on beliefs are quite limited. Often,  a single aggregated index is obtained by averaging the ordered survey responses.  As discussed in  \citet{pesaran2006survey},  the application of this method is limited to datasets with variables that are closely related and easily ordered. More concerning is that this method provides an aggregate summary without modeling any heterogeneity, and thus cannot explain  how beliefs differ.  Yet, \citet{manski04} and \citet{dmanski04}  document significant  heterogeneity in beliefs even for events where individuals are unlikely to have private information.  \citet{manski04}  suggests that the observed heterogeneity  can be related to differences in the way individuals process public information, but did not make precise what the process might be. We  suggest a model for this process. 

We model  heterogeneity in beliefs measured in surveys as coming from differences in  individuals' information choice.    An example helps fix ideas. Suppose that  we have survey responses for  individuals grouped by the majority political affiliation (Democrat or Republican) for the county  in which they live.  The goal is to understand the  heterogeneity in the survey responses given the observed group membership. Our economic model assumes  that individuals have access to many news sources of public information such as  CNN, ESPN, Discovery Channel,  Fox News, etc. and can choose one news source  to improve their naive belief about inflation. The model would predict that individuals   living  in a Republican-leaning county are more likely to (but do not always) choose Fox News, while those  living in a Democrat-leaning county  are more likely to (but do not always)  choose CNN. To the extent that Fox News provides more coverage encouraging fiscal conservatism and the dangers of monetizing the deficit,  this news source may leave its followers with a more pessimistic belief about future inflation.  The information effect induced by Fox News is combined with  an individual's naive expectation to determine whether the individual is optimistic or pessimistic about inflation.   In our economic model, an individual's observed group membership does not directly determine heterogeneous beliefs, but it influences the  source of  information that the individual chooses to process, which in turn influences their sentiments. 

Our economic model maps into a Bayesian hierarchical mixture model, also known as mixed membership model, that shares many similarities with Latent Dirichlet Allocation (LDA) of \citet{Blei2003} for text data.   Accordingly, we refer to our model as LDA-S. Mixed membership models are used in a variety of applications to cluster  high-dimensional discrete data, including text, network, and genomic data. 
Our contribution  is to show that a particular mixed-membership model  can be motivated from  an economic  model in which belief heterogeneity is linked to group-level differences in processing of common information, and that this model can guide Bayesian model estimation, identification, and parameter interpretation. LDA-S uses multinomial distributions  to explicitly take into account the categorical nature of the data. The hierarchical structure comes from the fact that the observed group membership  affects the observed outcome through the unobserved choice of information source. The econometrician specifies a prior distribution for how group membership affects the choice of the information sources and another prior for how information affects the responses to the survey. The  economic assumptions not only guide the choice of these  priors, but also facilitate interpretation of  the  posterior estimates. 

We derive two versions of the model: one  for  the static cross-section case, and one for the dynamic case when the belief formation process is time-dependent.   To illustrate the economic insight gained from estimating heterogeneous types in qualitative survey data, our first example estimates a dynamic LDA-S model from the Michigan data to show that the estimated proportions of belief types are linked to macroeconomic factors like uncertainty and unemployment; this information is not apparent when working only with the aggregate consumer sentiment index currently published.

In the second example, we show that the estimated belief types can be used as estimates of fixed effects in subsequent  regressions. We  re-examine the estimation of returns to education considered in \citet{Card1995}. Using data from the National Longitudinal Survey of Young Men, we first estimate belief types  using LDA-S. These  estimates are then used to control for unobserved heterogeneity and estimate heterogeneous effects in a second step OLS regression of income on education. We find an important source of heterogeneity in returns to education relating to  whether an  individual believes  `internal' versus `external' factors determine success.


\section{Related Econometrics Literature } 
There are three types of categorical variables: nominal variables which represent unordered categories (such as zip code and NAICS codes),   numerical (count) data, which represent ordered data such as age, and ordinal data (such as satisfaction metrics of excellent/good/bad) that are qualitative in nature. Ordinal data are said to be non-metrical when the distance between two categories has no interpretation. Survey data on categorical beliefs are often of this latter type.
 
There is a small existing literature on the methodology for analyzing qualitative survey data on beliefs. \citet{carlson1975inflation} uses a model with latent thresholds to derive aggregate probabilistic expectations from qualitative trichotomous responses on price expectations, assuming underlying quantitative inflation expectations are distributed normally. Structural modeling of categorical variables generally requires strong parametric assumptions  for  estimation of polychoric correlations, which is computationally burdensome and not easily scalable. For a discussion on these methods, see \citet{Ng2015}. Most often, a simple averaging method is used to aggregate qualitative beliefs data.  For example, the Gallup Poll Weekly Economic Confidence Index is derived from averaging responses to two questions on the current and future state of the economy, where ``good/excellent" or ``getting better" is assigned a numeric value of 1 and ``poor" or ``getting worse" is assigned a numeric value of -1. \citet{pesaran1985formation} uses a weighted average instead. Neither of these methods capture heterogeneity in responses. 

For analyzing  other types of categorical survey data, a popular method among social scientists is the  `asset index' method of \citet{filmer-pritchett:01,Filmer2001} used to summarize categorical data on asset ownership for  households in developing countries. The method converts all categorical variables to binary variables, and applies PCA to the transformed matrix of responses. The method is simple, non-parametric and requires nothing more than the ability to compute a singular value decomposition. PCA has also been applied to a matrix of responses directly when the outcomes are ordered, so, in principle, all PCA-based methods can be used to analyze qualitative beliefs as well.

Despite its simplicity, applying the Filmer-Pritchett  method to analyze categorical survey data has known drawbacks.   In addition to conceptual issues concerning what the indices actually measure, as discussed in   \citet{davidsonetal,wittenberg-leibbrandt} and references therein, there are also methodological issues.  Foremost is the problem  that averaging methods ignore the fact that the distance between different categories may not be constant. Furthermore,  converting multinomial outcomes to binary variables will introduce spurious negative correlations within the multiple columns that are mapped from a single question. These issues are discussed in  \citet{kolenikov2009, vyas-kumaranayake:06,lovatonetal}, among others. These limitations motivate our analysis.

The framework we adopt is a hierarchical Bayesian  latent class model. Latent class models were first introduced in \citet{lazarsfeld:50} to analyze dichotomous attributes based on a survey sample consisting of individuals who are assumed to belong to distinct classes. It was theorized in \citet{goodman:79} as a model for categorical variables. A multinomial latent class model for discrete data assumes that discrete survey data are generated by a finite mixture of multinomial distributions and that the mixture probabilities are the same for each individual. In contrast, the hierarchical latent class model  allows for heterogeneous mixture probabilities. Although latent variable models are widely used in economic analysis,  hierarchical latent variable models are far less common.  The analysis of \citet{dhfm} is for continuous data, while \citet{catania-mari:20} considers count data.

Hierarchical models for categorical data are known in the statistics literature as mixed membership models.   The first model in this class is the  grade-of-membership model   introduced in \citet{woodbury:78} that allows individual units to belong to multiple classes simultaneously.  A Bayesian version of it was used in  \citet{Erosheva02} to  analyze disability data. As seen from the collection of papers in \citet{handbook-mixedmembership}, many specifications and assumptions are possible within models in the mixed membership class. Particularly popular is   Latent Dirichlet Allocation (LDA) of  \citet{Blei2003} for modeling topics in documents, which is similar to  the admixture (mixture of mixture)  model   independently developed in  \citet{pritchard:00}.  \citet{erosheva2004mixed} develop a related mixed membership model where topics in scientific publications are associated with multiple multinomial distributions, one for words and one for references. \citet{Erosheva02} and \cite{Erosheva2007}  develop mixed-membership models as  a unifying framework for the soft clustering of categorical data  and show that  every mixed membership model has a finite mixture representation.  Applications include  social networks as in \citet{airoldi:08}, voting patterns and political beliefs as in \citet{gross-MV}, and genetic studies, such as associating genotypes with diseases as in \citet{falush:03}. \citet{draca-schwarz:20} use a mixed membership model to analyze political ideology. Their adaptations to economic analysis remain quite limited, though  LDA is getting  increased attention.  Recent applications include  the analysis of CEO time usage in \citet{bandiera2020ceo}, reducing the dimension of FOMC transcripts in \citet{Hansen2018}, and predicting consumer purchases in \citet{ruiz2020shopper}. Although there is a wide variety of mixed membership models available for various applications, it is not always clear to economists how to choose a model for a given situation and under what conditions the parameter estimates can be given an economic interpretation.

\section{An Econometric  Model of  Beliefs Based on Information Acquisition}
 The goal of the econometric exercise is to explain the heterogeneity in  survey responses using a model that (i) explicitly recognizes that many of the responses  are  in categorical  form,  (ii) allows the survey responses to have cross-section dependence; (iii)  connects unobserved heterogeneity in individuals to observed heterogeneity in characteristics and responses, and (iv) provides an economic interpretation to the unobserved heterogeneity. 

We will use  an economic model of information acquisition to motivate the econometric analysis. There is extensive theoretical work on the role of information in shaping economic agents' decisions. Costly information acquisition has been used as an amplifier of discrimination in the field experiment of \citet{bartovs2016attention}, and as a justification for `mistakes' in choices in the theoretical analysis of \citet{caplin2015revealed}. Empirical work has also used costly information acquisition as a possible  explanation for   heterogeneous responses  in the  Michigan Survey data. In \citet{carroll2003macroeconomic}, households occasionally update their inflation forecasts based on reports provided by the  Survey of Professional Forecasters. \citet{branch2004theory} models individuals as econometricians, and assumes that heterogeneous responses are generated by households choosing between a set of costly econometric prediction methods. These models are designed for univariate continuous outcomes and as a result do not make precise how information acquisition affects qualitative beliefs. 
 
\subsection{LDA-S for Cross-Section Data}

There are $N$ individuals indexed by $i$, and each individual is a  member of a group $d_i \in \mathbb{G}= \{1,\ldots, G\} $. There are $J$ discrete survey responses in the  dataset, where each question $j$ has $L^j$ possible responses. Individuals choose a single response $x_{ij} \in \mathbb{L}^j = \{1, \ldots, L^j\}$ for each question $j$ by choosing the option $v$ most appropriate for question $j$ as indicated by a  score function $B_{ij}(v) $, which depends on the set of information an individual has processed.

The information acquisition  model we consider is one of sequential choice and is motivated by  \citet{ruiz2020shopper} which recognizes that hierarchical models can be rationalized by economic models of myopic sequential choice.   An individual $i$ chooses which  of the $K$ sources of information, $z_{i} \in \mathbb{K} = \{1,\ldots,K\}$ to consume by maximizing their utility, which 
 incorporates a group  affinity for each information source $\bm u_{g,:} \in \mathbb{R}^{k}$  and an individual specific effect $e_{ik} \in \mathbb{R}$:
 
\begin{equation}  z_i = \arg \max_{k \in \{1,\ldots,K\}}  U_i(k) = \sum \limits_{j=1}^K\mathbbm{1}(k=j)(u_{d_i,j} + e_{ij} ) \label{eq:cinfo} \end{equation} 
The chosen source of information $z_i$  determines an individual's belief type. The observed heterogeneity of an individual's group membership is linked to the unobserved heterogeneity of an individual's belief type  by a function $\pi: \mathbb{G}  \rightarrow \Delta^{K-1}$  where 
 \[ \pi_{gk} = \mathbb P(z_{i}=k | d_i=g) = \mathbb P\left(u_{gk} + e_{ik}  = \max_{j\in \mathbb{K}} (u_{gj} + e_{ij}) \right). \] 
The probability that an individual $i$ selects information source $k$ is the probability that  $ u_{gk} - u_{gj} + e_{ik} - e_{ij} >=0$ for all $j \in K$.

The information source that is chosen influences the actual response that an individual makes. In the model, the actual response to survey question $j$ is optimal in the sense that it maximizes the individual's score function for each response. 

\[  x_{ij} = \arg \max_{v \in \{1, \ldots, L^J\}}  B_{ij}(v) = \sum \limits_{u=1}^{L_j} \mathbbm{1}(v = u) (q^j_{z_i,u} + s^j_{iu} ).\]

The score for each response $B_{ij}(v)$ has two components: 
\begin{enumerate}
\item An individual effect  $s^j_{iv} \in \mathbb{R}$ drawn independently for each $i$,$j$ and $v$ from distribution $S$. 
\item A type-specific information effect $\bm q^j_{k,:} \in \mathbb{R}^{L^j}$ drawn independently for each $k$ from some distribution $Q$. 
\end{enumerate} 

Here, $\bm s_{i,:}^j$  is interpreted as the naive score that the individual assigns to the question in the absence of any information.
 If the information is uninformative and $q^j_{kv} =0$ for all $v \in \{1,\ldots,L^j\}$, then the individual selects the naive response with the maximum $s^j_{iv}$.  In general, however, the acquired information that is shared by members of the same belief type will affect the response. 

The expected choices induced by each information source can be described by a map $\bm \beta^j: K  \rightarrow \Delta^{L^j-1}$. Precisely,  $\beta^j_{k,v}$ is a random variable that gives the probability that an individual with information source $z_i=k$ believes that option $v$ is the appropriate response to question $j$. 
\[ \beta^j_{kv} = \mathbb P(x_{ij} = v | z_i = k) = \mathbb P\left(q^j_{z_i,v} + s^j_{iv}  = \max_{u\in \mathbb{L}^j} (q^j_{z_i,u} + s^j_{iu}) \right).  \]

To complete the analysis, we   assume  that (a) $\bm u_{g,:}$ is independent over $g$ with distribution $ F_u^g$ and (b)  $e_{ik}$ is independent over $i$ and $k$ with distribution $ F_e$. The economic model has the  following  conditional independence properties: 

\begin{asml}
The source of information $z_i$ chosen by individual $i$ is independent of the source of information chosen by individual $h$ conditional on each individual's group membership and $\bm{\Pi}$, so that
$p(\bm{z} | \bm{d}, \bm \Pi) = \prod_{i=1}^{N} p(z_i | d_i , \pi_{d_i,z_i}) = \prod_{i=1}^N \pi_{d_i,z_i} $
\end{asml} 

\begin{asml} The group-specific mixtures $\bm \pi_{g,:}$  are independent across groups, so $p(\bm{\Pi}) = \prod_{g=1}^G p( \bm{\pi}_{g,:}) $.  
\end{asml} 


\begin{asml} Conditional on information source $\bm{z}$, $x_{ij}$ is independent of $x_{hr}$ for $j\ne r$ and $h \neq i$, and is independent of an individual's group membership $d_i$. $p(\bm{X} | \bm{d}, \bm{\Pi},\bm{z},\bm{\beta}) = \prod_{i=1}^N p(\bm{x}_{i,:} | z_{i},\bm{\beta})  = \prod_{i=1}^N \prod_{j=1}^J p(\bm{x}_{ij} | z_{i},\bm{\beta}) $. \end{asml}

\begin{asml} Conditional independence of belief types, so $\bm{\beta}^j_{k,:}\perp \bm{\beta}^j_{h,:}$ for $h\neq k$, and conditional independence of beliefs and mixtures so that $\bm{\beta}\perp \bm{\Pi}$. \end{asml} 

Property S1 follows from the two independence assumptions on $\bm u_{g,:}$ and $e_{ik}$. As  group-specific affinities for each information source are independent,  S2  rules out the selection patterns of one group influencing the access to information of a second group.
Since  all dependence across questions in individual responses are channeled through the common information source, S3  rules out individual preferences that are correlated across questions and unrelated to the information source. The terms in the score function $B_{ij}(v)$ are independent of the parameters that determine an individual's choice of information $z_i$, which leads to S4.  This completes the economic model.

The econometrician does not observe $\bm u$, $\bm e$, or their distributions ($F_u$ and $F_e$), so treats the $G\times K$ matrix  $\bm \Pi$ as  random. He proceeds with estimation by taking the features of the model S1-S4 as maintained assumptions. He specifies a  prior  on which $\bm{\Pi} \in \mathbb{G} \times \Delta^{K-1}$ are more likely through a Dirichlet distribution with hyperparameter $\bm \alpha_{g,:} \in \mathbb R^{K}$, which is the conjugate prior to the multinomial $\bm\pi_{g,:}$.  The econometrician also  does not know the distributions $S$ and $Q$, so specifies a prior as to what  sort of belief structures are more likely. Since each $\bm \beta^j_{k,:}$ is a multinomial distribution,  a Dirichlet prior with hyperparameter $\bm \eta^j_{k,:} \in \mathbb{R}^{L^j}$ is used:
\[ 
 \bm \pi_{g,:}  \sim \mbox{Dirichlet} (\bm \alpha_{g,:} ), \quad\quad
\bm  \beta^j_{k,:}   \sim \mbox{Dirichlet} (\bm \eta ^j_{k,:} ).  \] 
Thus, under the structure of the economic model,  we arrive at the statistical model for LDA-S 
 defined by the following equations: 
\paragraph{Model LDA-S}
 \begin{subequations}
\begin{align} 
 x_{ij} | \bm{\beta}, z_{i} &\sim \mbox{Multinomial} (\bm{\beta}_{j,z_{i}}), \label{eq:xij} \\
 z_{i} | \bm{\pi}_{d_i,:} &\sim \mbox{Multinomial}(\bm{\pi}_{d_i,:})  \label{eq:zg}\\
 \bm \pi_{d_i,:} &\sim \mbox{Dirichlet} (\bm \alpha_{d_i,:}) \label{eq:pi},  \\ 
\bm{\beta}^j_{z_i,:} &\sim \mbox{Dirichlet} (\bm{\eta}^j_{z_i,:}) \label{eq:beta}
\end{align} 
\end{subequations}
where  individuals are indexed by $i=1,\ldots,N$ and categorical responses  in the survey are indexed by $j=1,\ldots,J.$ Using $p(x_{ij}=v |\bm{\beta},z_{i})= \beta^j_{z_{i},v }$ and $p(z_i=k|\bm{\pi}_{d_i,:})=\pi_{d_{i},k}$, we can write the  joint distribution of the model as
\[
p(\bm{\beta},\bm{\Pi},\bm{z}, \bm d , \bm{X}) = \prod_{j=1}^J \prod_{k=1}^K p(\bm{\beta}^j_{k,:}) \prod_{g=1}^G p(\bm{\pi}_{g,:}) \prod_{i=1}^{N} \pi_{d_i,z_{i}} \prod_{j=1}^J \beta^j_{z_{i},x_{ij}}.
\]

\subsection{Dynamic LDA-S} 
 Many surveys on beliefs are conducted repeatedly, each time with a possibly different sample of individuals.  The Michigan Survey of Consumers  is perhaps the best known example as it has been in existence for over 50 years.  The Bank of Canada interviews the business leaders of about 100 firms to gather perspectives on business outlook, while the Bank of England solicits attitudes towards inflation. 

As in the static model, we will motivate with an example,  one that attributes differences in economic sentiment to time of response and the outlook of  business news. For the dynamic example, we group individuals by the time at which they respond to the survey, rather than the location in which they live. In each month, there are a variety of news articles that convey different sentiments, but only a few are relevant to the survey questions.  For example,  some articles might have an optimistic tone and others may have a pessimistic one. The model postulates that the individual survey response  will depend on the prevalence of the article type at time $t$ (the time specific effect), as well as a random idiosyncratic preference for the  article type (the individual specific effect). In the dynamic model, $\pi_t$ varies with time to allow  the proportion of optimistic and pessimistic information about the economy to vary over time; during bad times, it is easier to access pessimistic news and incorporate that information into beliefs. An individual's selection of news type determines their belief type, and is labeled $z_i$. An individual who reads negative news on declining energy prices and bankrupt oil companies is more likely to respond to a survey indicating that they believe the economy is weak. An individual who selects the positive news source describing a booming tech industry with new product offerings and rising stock prices is more likely to respond to a survey indicating that they believe the economy is robust. This is represented by $\bm \beta^j $, which gives the expected responses of individuals who have absorbed different types of news sources.

We will continue to work with a $N \times J$ matrix of survey response data but will refer to the dynamic model as LDA-DS, denoting
 the time at which an individual responds to the survey by $s_i \in \{1,\ldots,T\}$.  The choice described in Equation \ref{eq:cinfo} is now: 

\begin{equation} z_i = \arg \max_{k \in \{1,\ldots,K\}}  U_i(k) = \sum \limits_{j=1}^K\mathbbm{1}(k=j)(u_{s_i,j} + e_{ij} )  \end{equation} 
In the static  model,  $u_{t,j}$ is assumed to be independent of $u_{s,j}$ for any $t \neq s$. But, $u_{t,j}$ is meant to capture the net cost and benefit of accessing news type $j$ for individuals who form beliefs at time $t$. Economic conditions tend to be persistent, so a high proportion of negative news in month $t-1$ is  likely followed by pessimism  in month $t$. Thus, we now assume  time dependence in the proportions of individuals of each belief type. As a consequence, while features S1, S3 and S4 of dynamic model are the same as in LDA-S (and now labeled DS1, DS3 and DS4),  S2 needs to be replaced with  the following: 

\paragraph{Assumption DS2:} Mixture proportion $\bm{\pi}_{t,:}$ is a first order Markov process. $\bm{\pi}_{t,:}$ is independent of $\bm{\pi}_{t-s,:}$ conditional on $\bm{\pi}_{t-1,:}$ for $s >1$.  Let
$ \pi_{tk}= \exp(\tilde{\pi}_{tk})/\sum \limits_{k=1}^K \exp( \tilde{\pi}_{tk} ) $ be lognormally distributed and  satisfy $\sum^K_k\pi_{gk}=1$, where  for $k=1,\ldots,K$ and $t=1,\ldots,T$,
\[ \tilde\pi_{tk}=\tilde\pi_{t-1,k}+ w_t, \quad w_t\sim N(0,\sigma^2_k).\]


 The econometrician now takes  DS1 - DS5 as maintained assumptions. For $i=1,\ldots,N$ and $j=1,\ldots,J$, the model is characterized by
\begin{subequations}
\begin{eqnarray}
   x_{ij} | \bm{\beta}, z_{i} &\sim& \mbox{Multinomial}(\bm \beta^j_{z_{i},:}), \label{eq:de1} \\
 z_{i}|\bm{\pi}_{s_i,:} &\sim& \mbox{Multinomial}(\bm\pi_{s_i,:}), \label{eq:de2}\\
 \tilde{\bm{\pi}}_{s_i,:} | \tilde{\bm{\pi}}_{s_i-1,:},\sigma^2_{z_i}&\sim& \mbox{Normal}(\tilde{\bm{\pi}}_{s_i,:},\sigma^2_{z_i}), \label{eq:de3} \\
 \sigma^2_{z_i} &\sim& \mbox{InverseGamma}(v_0,s_0), \label{eq:de4} \\
 \bm{\beta}^j_{z_i,:} &\sim& \mbox{Dirichlet}(\bm{\eta}^j_{z_i,:}) \label{eq:de5}.
\end{eqnarray}
\end{subequations}
The joint probability distribution of the dynamic hierarchical latent class model becomes
\[ 
p(\bm{\beta},\bm{\Pi},\bm{z},\bm{X}) = \prod_{j=1}^J\prod_{k=1}^K p(\bm{\beta}^j_{k,:}) p(\sigma^2_k) \prod_{t=1}^T p(\tilde{\bm{\pi}}_{t,:} | \tilde{\bm{\pi}}_{t-1,:}) \prod_{i=1}^N p(z_i|\bm{\pi}_{s_i,:}) p(x_{ij}|z_i,\bm{\beta}).
\]

\subsection{Relationship to Latent Class, Mixed Membership Models and NMF}

We now describe how LDA-S and LDA-DS are related to existing  models in both statistical and economic terms. Consider first a non-hierarchical Bayesian latent class (LC) model. To justify the absence of a group-specific effect would require that  $\bm u_{d_i,:}$  is the same  for each individual.  Under such a restriction, any observable heterogeneity $d_i$ is independent of an individual's choice of information $z_i$. This, however, would be inconsistent with the  substantial evidence that heterogeneity in responses is related to heterogeneity in observed characteristics, such as group membership. For example, \citet{manski04}  finds  gender and schooling type to be  correlated with an individual's optimism. Allowing  an individual's choice of information to depend on observed characteristic $d_i$ endows LDA-S with more flexibility. 

The GoM (Grade of Membership) model of \citet{Erosheva02} is a mixed membership model, which has been applied to survey data on disabilities, health, and political ideology. The difference between GoM and LDA-S is that for GoM, the type assignment parameter $\bm Z$ is a $N \times J$ matrix, but for LDA-S, $\bm z$ is a  $N \times 1$ vector. This means that in GoM, the response to each  question is assigned its own class label, so each individual is a mixture over belief types. In both our economic model and LDA-S, we only have  an assignment to a single belief type.  This does not mean that a more sophisticated economic model cannot be written per se, but GoM cannot immediately be motivated by our model of information acquisition. Furthermore, the flexibility of GoM comes at a cost of many more parameters, and, in our experience, as further discussed  in Section 4, GoM is more susceptible to the problem of non-uniqueness for a given prior and dataset.  LDA-S  allows information source selection probabilities to co-move yet also vary across individuals. Assuming that each individual is a member of only a single belief type  but groups of individuals are mixtures over belief types simplifies interpretation and identification. 

The model in the mixed-membership class closest to ours is  LDA for estimating topics in a corpus of documents.    In topics modeling, the singular value decomposition of a word-document frequency matrix $\bm Y_D$ is known as Latent Semantic Indexing (LSI). A probabilistic variation of it, known as pLSI, was developed in \citet{hofmann:99} and \citet{hofmann:01}. pLSI treats the document-specific mixture over topics as a fixed parameter and the documents as a fixed collection. LDA is the Bayesian version of pLSI, which treats document-specific mixtures over topics as random, as specified by a Dirichlet distribution.   The appeal of  LDA over pLSI is that documents which have not been observed can be analyzed.  

LDA-S is an adaptation of LDA to survey response data. Instead of the frequency of word occurrences in documents, we analyze frequency of responses to questions in grouped individuals. In LDA, documents are modeled as mixtures over topics, which each involve different distribution over words. In LDA-S, we model survey responses as group-specific mixtures over $K$ belief types, each characterized by multinomial distributions over survey responses. A topic in LDA involves a single distribution over all words in the corpus. For each word $w_i$, the outcome variable is simply the identity of that word, so $J=1$. The joint likelihood for LDA can be factorized: 
\[
p(\bm{\beta},\bm{\Pi},\bm{z},\bm{w}, \bm{g}) = \prod_{k=1}^K p(\bm{\beta}_{k,:}) \prod_{d=1}^D p(\bm{\pi}_{d,:}) \prod_{i=1}^{N} p(z_{i}|\bm{\pi}_{g_i,:}) p(w_{i}|z_{i},\bm{\beta}).
\]  

By contrast, LDA-S sets $J\geq 1$ because in surveys we have $J$ survey questions and for individual $i$ we observe $x_{ij}$ for $j=1,\ldots J$. The belief types in the LDA-S model each specify $J$ multinomial distributions, one for each question response recorded in the columns of $\bm{x}_{i,:}$. So we may think of LDA-S as a multivariate variant of LDA with $J\ge 1$, but a simplified version of GoM since our $\bm z$ is an $N\times 1$ vector instead of a $N\times J$ matrix.  A concise comparison of LDA-S to  LDA is as follows: 
{\small
\begin{center}
  \begin{tabular}{l|l|ll}
    & LDA-S for types in responses & LDA for topics in documents\\ \hline
Data/outcome & $x_{ij}$ response of $i$ to question $j$ & $w_{i}$ word in corpus \\
Outcome dimension & $x_{ij} \in \{1,\ldots,L_j\}$ & $w_i \in \{ 1,\ldots,V\}$  \\ 
Outcomes per unit & $J \geq 1$ responses in $x_{i,:}$ & $J=1$ word index of $w_i$ \\ 

$N$ & \# individuals in survey & \# words appearing in corpus \\

Frequency matrix & $\bm Y_S$ (group-response) &  $\bm Y_D$ (document-word) \\
Latent size & $K$ information sources & $K$ topics \\
Mixture size &$G$, number of groups & $D$, number of documents \\ 
Class assignment & $z_i$ information source/belief type & $z_i$ assignment of word $i$ to topic \\
Membership & $d_i$ membership of individual $i$ in group & $g_i$ membership of word $i$ in doc \\ 
 Mixture & $\pi_{g,:}$ group-specific mixture over types & $\pi_{d,:}$ document specific mixture over topics \\
Outcome distribution & $\bm \beta^j_{k,:} $ for $x_{ij}$ with $z_i=k$  & $\bm{\beta}_{k,:}$ for $w_i$ with $z_i=k$\\

\hline
  \end{tabular}
\end{center}
}
\bigskip

The dynamic version of LDA introduced by \citet{Blei2006} incorporates dynamics into class proportions $\bm{\Pi}$ but also allows the topic distributions $\bm{\beta}$ to evolve according to different documents that are observed over time. Modeling time dependence in $\bm \beta$ would   require  the expected responses induced by each information source to change over time. While this is possible, it seems more appropriate to  explain variations in survey responses  by the variation in the proportion of belief types over time, while holding the beliefs for each type fixed.


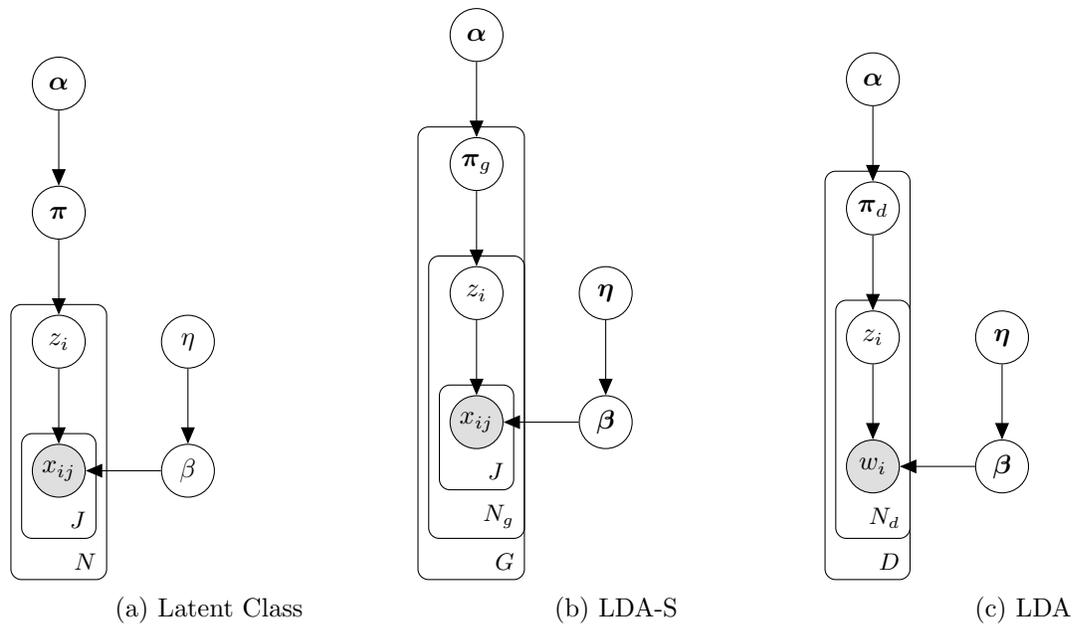
\begin{figure}
\begin{center}
\caption{Probabilistic Graphs for Hierarchical Latent Class Models \label{lcagraph}} 

\begin{subfigure}[b] {0.32\textwidth} 
\begin{tikzpicture}
 \node[obs]            (x) {$x_{ij}$};
 \node[latent, above=of x] (z) {$z_{i}$};
 \node[latent, above=of z] (g) {$\bm{\pi}$};
 \node[latent, right=of x] (b) {$\beta$}; 
 \node[latent, above=of b] (e) {$\eta$}; 
 \node[latent, above=of g] (a) {$\bm{\alpha}$};
 \edge {z} {x} ; %
 \edge{g}{z}; 
 \edge{b}{x}; 
 \edge{e}{b} 
 \edge{a}{g} 
  \plate{x}{(x)}{$J$}; 
 \plate {zx} {(z)(x)(x.north west)(x.south west)} {$N$} ;
\end{tikzpicture}
\subcaption{Latent Class} 
\end{subfigure}
\begin{subfigure}[b] {0.32\textwidth} 
\begin{tikzpicture}
 \node[obs]            (x) {$x_{ij}$};LDA-S \node[latent, above=of x] (z) {$z_{i}$};
 \node[latent, above=of z] (g) {$\bm{\pi}_g$};
 \node[latent, right=of x] (b) {$\bm{\beta}$}; 
 \node[latent, above=of b] (e) {$\bm{\eta}$}; 
 \node[latent, above=of g] (a) {$\bm{\alpha}$};
 \edge {z} {x} ; %
 \edge{g}{z}; 
 \edge{b}{x}; 
 \edge{e}{b} 
 \edge{a}{g} 
  \plate{x}{(x)}{$J$}; 
 \plate {zx} {(z)(x)(x.north west)(x.south west)} {$N_g$} ;
  \plate {gzx} {(g)(z)(x)(zx.north west)(zx.south west)} {$G$} ;
\end{tikzpicture}
\subcaption{LDA-S} 
\end{subfigure}
\begin{subfigure}[b] {0.32\textwidth} 
\begin{tikzpicture}
 \node[obs]            (x) {$w_{i}$};
 \node[latent, above=of x] (z) {$z_{i}$};
 \node[latent, above=of z] (g) {$\bm{\pi}_d$};
 \node[latent, right=of x] (b) {$\bm{\beta}$}; 
 \node[latent, above=of b] (e) {$\bm{\eta}$}; 
 \node[latent, above=of g] (a) {$\bm{\alpha}$};
 \edge {z} {x} ; %
 \edge{g}{z}; 
 \edge{b}{x}; 
 \edge{e}{b} 
 \edge{a}{g} 
 \plate {zx} {(z)(x) }{$N_d$} ;
  \plate {gzx} {(g)(z)(x)(zx.north west)(zx.south west)} {$D$} ;
\end{tikzpicture}
\subcaption{LDA} 
\end{subfigure}
\vspace*{.75in}

\begin{subfigure} [b]{0.3\textwidth} 
\begin{tikzpicture}
 \node[obs]            (x) {$x_{ij}$};
 \node[latent, above=of x] (z) {$z_{i}$};
 \node[latent, above=of z] (g) {$\bm{\pi}_i$};
 \node[latent, right=of x] (b) {$\bm{\beta}$}; 
 \node[latent, above=of b] (e) {$\bm{\eta}$}; 
 \node[latent, above=of g] (a) {$\bm{\alpha}$};
 \edge {z} {x} ; %
 \edge{g}{z}; 
 \edge{b}{x}; 
 \edge{e}{b} 
 \edge{a}{g} 
  \plate{x}{(x)}{$J$}; 
 \plate {zxg} {(z)(x)(g)(x.north west)(x.south west)} {$N$} ;
\end{tikzpicture}
\subcaption{GoM} 
\end{subfigure}
\begin{subfigure}[b] {0.4\textwidth} 
\begin{tikzpicture}
  \node[latent, above=of g] (a) {$\bm{\alpha}$};
 \node[obs]            (x) {$x_{ij}$};
  \node[latent, right=of x] (b) {$\bm{\beta}$}; 
 \node[latent, below=of b] (e) {$\bm{\eta}$}; 
 \node[latent, above=of x] (z) {$z_{i}$};
 \node[latent, above=of z] (g) {$\bm{\pi}_t$};
   \node[obs, right=of b] (x1) {$x_{ij}$};
     \node[latent, above=of x1] (z1) {$z_{i}$};
  \node[latent, above=of z1] (g1) {$\bm{\pi}_{t+1}$};
  
   \node[draw=none,right =of g1] (d1){$\hdots$}; 
   \node[draw=none,right=of x1] (d2){$\hdots$}; 
 
 \edge {z} {x} ; %
 \edge{g}{z}; 
 \edge{b}{x}; 
 \edge{e}{b} 
 \edge{a}{g} 
  \edge {z1} {x1} ; %
 \edge{g1}{z1}; 
 \edge{g}{g1} 
 \edge{b}{x1}; 
 \edge{a}{g1} 
 \edge{g1}{d1}
 
  \plate{x}{(x)}{$J$}; 
 \plate {zx} {(z)(x)(x.north west)(x.south west)} {$N_t$} ;
   \plate{x1}{(x1)}{$J$}; 
 \plate {z1x1} {(z1)(x1)(x1.north west)(x1.south west)} {$N_t$} ;
\end{tikzpicture}
\subcaption{LDA-DS} 
\end{subfigure}
\end{center}
\end{figure} 
Figure \ref{lcagraph} presents a graphical view that concisely summarizes the differences in statistical conditional independence assumptions across models. Figure (a) shows the standard LC model. Figure (b) shows that LDA-S is the same as an LC model, except that (i) $z_i$ is now a descendant of $\bm{\pi}_d$ rather than a common $\bm{\pi}$, which means $z_i$ is now independent of $z_j$ upon conditioning on a group-specific mixture, rather than an aggregate mixture $\bm\pi$. These changes from the basic model allow for more heterogeneity in the patterns of class assignment within the population. Comparing GoM in (d) to LDA-S in (b), the plate surrounding a row of $\bm{\Pi}$ is now $N$-dimensional, and the $J$-subscripted plate now incorporates both $x_{ij}$ and $z_{ij}$ since $\bm{Z}$ is now $N \times J$. Observe that in the LDA-S model, there is no link between $\bm{\pi}_g$ and $\bm{\pi}_h$ except through their common prior. But in LDA-DS there is a direct link between $\bm{\pi}_s$ and $\bm{\pi}_{s+1}$. The rest of the model remains the same, so we assume that the time-specific mixture over belief types captures all the time dependence. 

We close this section with two additional remarks.  First, note that $\bm{x}_{i,:}$ in LDA-S in (b) is $J$-dimensional, but $w_i$ in LDA in (c) is not. This is not just a modeling choice but is in some sense necessary. Modeling the $J$ responses for each individual as one outcome would be necessary if we assumed there were correlation in individual responses across questions beyond what is induced by an individual's belief type. However, this would require dealing with the joint distribution of responses for each individual, which would then be a multinomial distribution over all possible response permutations for a survey. For $J$ that is larger than a handful of questions, this is too high-dimensional.

Second, there are other  interpretations of LDA that will also help in understanding LDA-S. \citet{buntine:02,buntine-jakulin} show that LDA is a type of independent component analysis, while \citet{canny:04} shows that LDA is a type of factor analysis for discrete data. 
When the latent class assignments $z_i$ are integrated out, LDA can also be seen as factorizing the $D \times V$ (document-term) matrix $\bm Y_D$ into two low rank matrices, one representing the probability of topics given document, and one representing probability of words given topics. In lieu of \textsc{svd}, it is also possible to obtain a non-negative matrix factorization (NMF) of $\bm Y_D=\bm{ W}_D\bm { H}_D$ into two rank $K$ matrices, $\bm { W}_D$ and $\bm H_D$, both with non-negative entries. \citet{ding-etal:06} show that NMF and PLSI solve the same objective function. In LDA-S, the discrete data $\bm X$ is mapped into a (non-negative) frequency response matrix $\bm Y_S=(Y_{S1},\ldots, Y_{SJ})$  of dimension $G\times L$, $L= \sum_{j=1}^ J L_j $. In our analysis, belief type $z_i$ is itself of interest, so we do not integrate it out. But if this were done, the frequency matrix $\bm Y_{S}$, also known as contingency table, can also be seen as a product of two low rank matrices: $\bm Y_S=\bm H_S\bm W_S$.  The matrix $\bm H_S$ represents the probability of responses given belief type assignment and can be compared to a flattened version of the 3-dimensional matrix $\bm \beta$, while $\bm W_S$ is the probability of belief type assignment for each group. Though the NMF approach is non-parametric and does not specify a probability distribution for the latent variables,  the low rank factorization perspective is helpful in understanding the issues to be discussed in Section 4.

\section{Implementation Issues} 

 This section discusses three implementation issues:  estimation of the model, the choice of the number of belief types $K$, and identification for a given number of types $K$. 

\subsection{Sampling Algorithms}
An appeal  of LDA is that the hierarchical structure  makes the model interpretable and computationally feasible, and LDA-S inherits these appealing features.  We can also exploit computational tools developed for estimating LDA, which is a heavily researched topic in the last two decades. LDA can be estimated using either MCMC or variational inference methods. The latter approximates posterior distributions using optimization techniques and is reviewed in \citet{bleivariational}. 

For this paper, we will use MCMC methods, which are familiar to economists. In the static case, conjugacy of the Dirichlet distribution makes deriving a Gibbs sampler for posterior estimation quite simple. In each step of a Gibbs sampler, each variable is sampled from its conditional distribution, conditional on all other variables in the model.

\begin{enumerate} 
\item Sample $z_i$ conditional on $\bm{x}_{i,:}$,  $\bm{\beta}$, and $\bm \pi_{d_i,:} $. The conditional distribution of $z_i$ is multinomial: 
\[ p(z_i =k | \bm{x}_{i,:}, \bm{\beta}, \bm \pi_{d_i,:}) \propto \pi_{d_i,k} \prod_{j=1}^J \beta^j_{k,x_{ij}}, \quad i=1,\ldots N. \] 
\item Sample $\bm{\beta}$ conditional on $\bm{\eta}$, $\bm{X}$, and $\bm{z}$. The conditional distribution of $\bm{\beta}^j_{k,:}$ is Dirichlet, with: 
\[ \bm{\beta}^j_{k,:} \sim \mbox{Dirichlet}(\eta^j_{k,1} + C^{resp}_{jk1}, \ldots, \eta^j_{k,L_j} + C^{resp}_{jkL_j} ),\quad  j=1,\ldots,J, k=1,\ldots,K, \] 
where $C^{resp}_{jk\ell} = \sum \limits_{i=1}^N \sum \limits_{\ell=1}^{L_j} \mathbbm{1} (z_i=k) \mathbbm{1}(x_{ij} = \ell)$ 
\item Sample $\pi_{g,:}$ conditional on $\bm{\alpha}$, $\bm d$, and $\bm{z}$. The conditional distribution of $\bm{\pi}_{g,:}$ is Dirichlet, with: \[ \bm{\pi}_{g,:} | \bm{\alpha}, \bm{z} \sim \mbox{Dirichlet}(\alpha_{g1} + C^{grp}_{g1} , \ldots, \alpha_{gK} + C^{grp}_{gK} ), \quad g=1,\ldots,D, \]  where $C^{grp}_{gk} = \sum \limits_{i=1}^N \mathbbm{1}(z_i=k)\mathbbm{1}(d_i=g)$.
\end{enumerate} 

Steps 1 and 2 of the Dynamic LDA-S sampler are the same as in LDA-S. But Step 3 is more complicated
because the dynamic model does not have a conjugate prior distribution for the state mixtures $\bm{\pi}_{t,:}$. \citet{Blei2006} specify a lognormal formulation for $\pi_{tk}$ and use a variational Kalman Filter for estimation. Using a Metropolis-Hastings step to sample from the posterior of the lognormal distribution for $\bm \pi_{t,:}$ would involve slow exploration of the posterior and is difficult to scale to datasets with a large number of respondents or time periods. \citet{Bradley2018} and \citet{Linderman2015} replace the lognormal formulation to improve sampling speed and convergence. 

Rather than changing the formulation for the dynamics, we adapt recent advances in MCMC methods to improve sampling speed and convergence. We use a sampling approach similar to the one in \citet{Bhadury2016}, which is a method known as Stochastic Gradient Langevin Dynamics (SGLD) developed in \citet{Welling2010} for learning Bayesian models from large scale datasets. In brief, SGLD works as follows. For parameter of interest $\theta$, prior $p(\theta)$, and data $(x_1,\ldots, x_N)$, let the likelihood be $L (\theta)$. Instead of using the full sample gradient $ g(\theta) =\nabla \log p (\theta) + \sum_{i=1}^N \nabla \log L (x_i|\theta)$
 of the log posterior distribution to find the mode, 
 SGLD updates $\theta$ at step $r$ according to
\begin{eqnarray*}
   \Delta \theta^{(r)}
   &=&\frac{\epsilon^{(r)}}{2} \bigg( g(\theta^{(r)}) + h^{(r)}(\theta^{(r)})\bigg)+\psi^{(r)} \quad \quad \psi^{(r)}\sim N(0,\epsilon^{(r)})\\
   &=& \frac{\epsilon^{(r)}}{2} \bigg( \nabla \log p(\theta^{(r)}) + 
   \frac{N}{M} \nabla \log L(x_i^{(r)}|\theta^{(r)}) \bigg) + \psi^{(r)}.
\end{eqnarray*} 
where $h^{(r)}(\theta)= \nabla \log p (\theta)+ \frac{N}{M} \sum_{i=1}^n \nabla \log L (x^{(r)}_{i}|\theta)-g(\theta)$
and $\epsilon^{(r)} $ is the stepsize. 
The acronym SGLD comes from the fact that it (i) uses {\em mini-batches} of the data of size $M\leq N$ (the stochastic gradient part), and (ii) adds noise (the Langevin dynamics part). The SGLD method ensures that the parameters reach the maximum of the posterior distribution quickly, and once it reaches the MAP value, the sampling procedure enters a Langevin dynamics stage, where the noise added to the parameter updates result in exploration of the posterior distribution. Though the above defines a non-stationary Markov chain, \citet{Welling2010} show that $\theta^{(r)}$ will converge to samples from the true posterior distribution with careful choice of the stepsize $\epsilon$. In particular, $\epsilon^{(k)}$ must satisfy $\sum_{k=1}^\infty \epsilon^{(k)}=\infty$ and $\sum_{k=1}^\infty (\epsilon^{(k)})^2 < \infty$.  The MCMC draws not only give the posterior mode, but the entire posterior distribution.

To use this approach in our setting requires making precise the density of interest and its derivative. Let $\phi(\cdot)$ be the standard normal density function. The condition distribution of $\tilde{\pi}_{t+1}$ is proportional to: 
\[ p(\tilde{\bm{\pi}}_{t,:} | \tilde{\bm{\pi}}_{t-1,:}, \tilde{\bm{\pi}}_{t+1,:}, z_i) \propto \prod_{k=1}^K \phi \left(\frac{\tilde{\pi}_{t,k} - \tilde{\pi}_{t-1,k}}{ \sigma^2_k} \right) \phi \left(\frac{\tilde{\pi}_{t+1,k} - \tilde{\pi}_{t,k}}{\sigma^2_k}\right) \prod_{i=1}^{N} \pi_{s_i,z_i} ^{\mathbbm{1}(s_i=t)}  \]

Then, the SGLD step for $\tilde{\pi}_{tk}$ is a combination of gradient descent towards the maximum of the posterior distribution and additional noise that ensures the full posterior distribution is explored. Our approach nonetheless differs from SGLD in one way. In the original paper, the gradient descent step is taken with respect to a random subsample of data (ie. the mini-batch). For many survey datasets and certainly applications considered in this paper, the number of respondents is small enough that we do not need to take subsamples when updating the parameters. Hence we use the batch size that equals the size of the dataset. In step $r$ of the Gibbs Sampler, for each $k=1, \ldots, K$,
\[ \tilde{\pi}^{(r)}_{t,k} - \tilde{\pi}^{(r-1)}_{t,k} = \frac{\epsilon_r}{2} \frac{\partial \log p(\tilde{\bm{\pi}}^{(r-1)}_{t,:} | \tilde{\bm{\pi}}^{(r)}_{t-1,:}, \tilde{\bm{\pi}}^{(r-1)}_{t+1,:}, z^{(r-1)}_i)}{\partial \tilde{\pi}_{t,k}} + \psi_i, \qquad \psi_i \sim N(0,\epsilon_r) \]
The first derivative is easily computed as follows: 
\[ \frac{\partial \log p(\tilde{\bm{\pi}}_{t,:} | \tilde{\bm{\pi}}_{t-1,:}, \tilde{\bm{\pi}}_{t+1,:}, z_i)}{\partial \tilde{\pi}_{t,k}} = \frac{-1}{\sigma^2_k}(\tilde{\pi}_{k,t} - \tilde{\pi}_{k,t-1}) - \frac{1}{\sigma^2_k} (\tilde{\pi}_{k,t+1} - \tilde{\pi}_{kt}) + n_{tk} - N_t \pi_{k,t} \]
where $N_t = \sum \limits_{i=1}^N \mathbbm{1}(s_i=t)$ and 
 $ n_{tk} = \sum_{i=1}^{N} \mathbbm{1} (z_{t} =k) \mathbbm{1} (s_{i} =t) $. 

 We follow the literature and use a step size of $\epsilon_r = a(b+r)^{-c}$ at step $r$ with $a=0.01$, $b=1$ and $c=0.5$. Tuning these parameters to get fast convergence and good posterior exploration can be challenging. \citet{Welling2010} show that as $\epsilon_r$ decreases, the acceptance probability approaches 1, so that a Metropolis-Hastings test is unnecessary as the acceptance probability is close to 1. The last part of the sampler for LDA-DS involves $\sigma^2_k$, which, conditional on all the other variables in the model, has an Inverse Gamma distribution. Let $v_1 = v_0 + T$ and $s_{1k}= s_0 + \sum \limits_{t=1}^T(\tilde{\pi}_{t,k} - \tilde{\pi}_{t-1,k})^2 $. Then, 
$\sigma_k | \bm{\Pi} \sim \mbox{IGamma}(v_1,s_{1k})$.

 \subsection{Identification and Choice of Priors}
 Finite mixture models are susceptible to weak and under-identification, and the researcher must also be alert to label switching as the ordering of the estimated latent class is arbitrary. See \citet{jasra-holmes-stephens} and \cite{masyn:13} for a recent review of the literature. Since
 LDA-S is a variation of LDA, and LDA is a finite mixture model, we begin with a discussion of identification issues concerning LDA.    As noted earlier, LDA can be seen as a problem of finding a non-negative factorization of the matrix $\bm Y_D$, but such a matrix factorization is not unique. Some applications of LDA do not require uniqueness, such as prediction of the content of new documents. It is nonetheless still useful to understand the two sources of non-uniqueness and how to deal with the problem. First, if a solution satisfies $\bm Y_D= \bm W_D \bm H_D$, then any positive definite $\bm Q$ producing $\tilde{\bm W}_D=\bm W_D\bm Q_D\ge 0$ and $\tilde{\bm H}_D=\bm Q_D^{-1}\bm H_D \ge 0$ is also solution. For example, $\bm Q_D$ can be the permutation of a diagonal matrix with positive diagonal elements, so scaling and permuting the rows and columns of $\bm W_D$ and $\bm H_D$ can produce equivalent solutions. For uniqueness of the non-negative matrix factorization problem, see \citet{gillis:04}, \citet{brie:05} \citet{hoyer:04}, \citet{huang-etal:14}.  More problematic is that there may be additional $\tilde{\bm W}_D$ and $\tilde{\bm H}_D$ matrices not obtained from rotations by $\bm Q_D$. Imposing summing up constraints on the rows of $\bm{W}_D$ and $\bm H_D$ are not usually  enough to tie down a unique LDA solution, so side conditions are needed. This is not surprising because the LDA is a mixture model known to be non-identifiable without further assumptions. The Bayesian approach is to impose non-exchangeable priors. For Dirichlet priors, \citet{griffiths-steyvers:07} show that the posterior distributions of $\bm \beta$ and $\bm \pi$ in LDA have a mode of: 
 \[\beta_{kv}=\frac{C_{kv}^{\text{word}}+ \eta_{v}}{\sum_{v=1}^V C_{dk}^{\text{word}}+\sum \limits_{v}^V \eta_v}, \quad \quad\quad
 \pi_{d k}=\frac{C_{d k}^{\text{doc}}+\alpha_k}{\sum_{k=1}^ K C_{dk}^{\text{doc}}+\sum \limits_{k}^K \alpha_k}.
 \]
 where $C^{\text{word}}$ is a $K \times V$ matrix of word counts for each topic, and $C^{\text{doc}}$  is a $D \times K$ matrix of document counts for each topic. From this representation, it is clear that the prior only shrinks towards but does not restrict $\bm \beta$ or $\bm \pi$ to any particular value. See also \citet{kmon:19} for a discussion of how the choice of  priors influence identification in LDA.

 Like LDA, the LDA-S decomposition is not unique. Following the literature, we impose non-exchangeable priors to ensure that the same belief type is identified in each MCMC draw. Specifically, we assign  a Dirichlet prior for the expected information choice conditional on group membership  $p(\bm \Pi)$,  and another Dirichlet prior  for expected response choice conditional on information choice  $p(\bm \beta)$.  The posterior expected values for $\bm \beta$ and $ \bm \Pi$ that emerge from  steps 2 and 3 of the Gibbs Sampler are analogous to the two expressions for LDA just presented, with $C^{\text{word}}$ replaced by counts for the responses for a question and $C^{\text{doc}}$ replaced by counts for each group.  

\begin{figure}[ht]
\caption{Density of Beta($\alpha$, $\alpha$) Distribution } 
\centering
\includegraphics[width=0.5\textwidth]{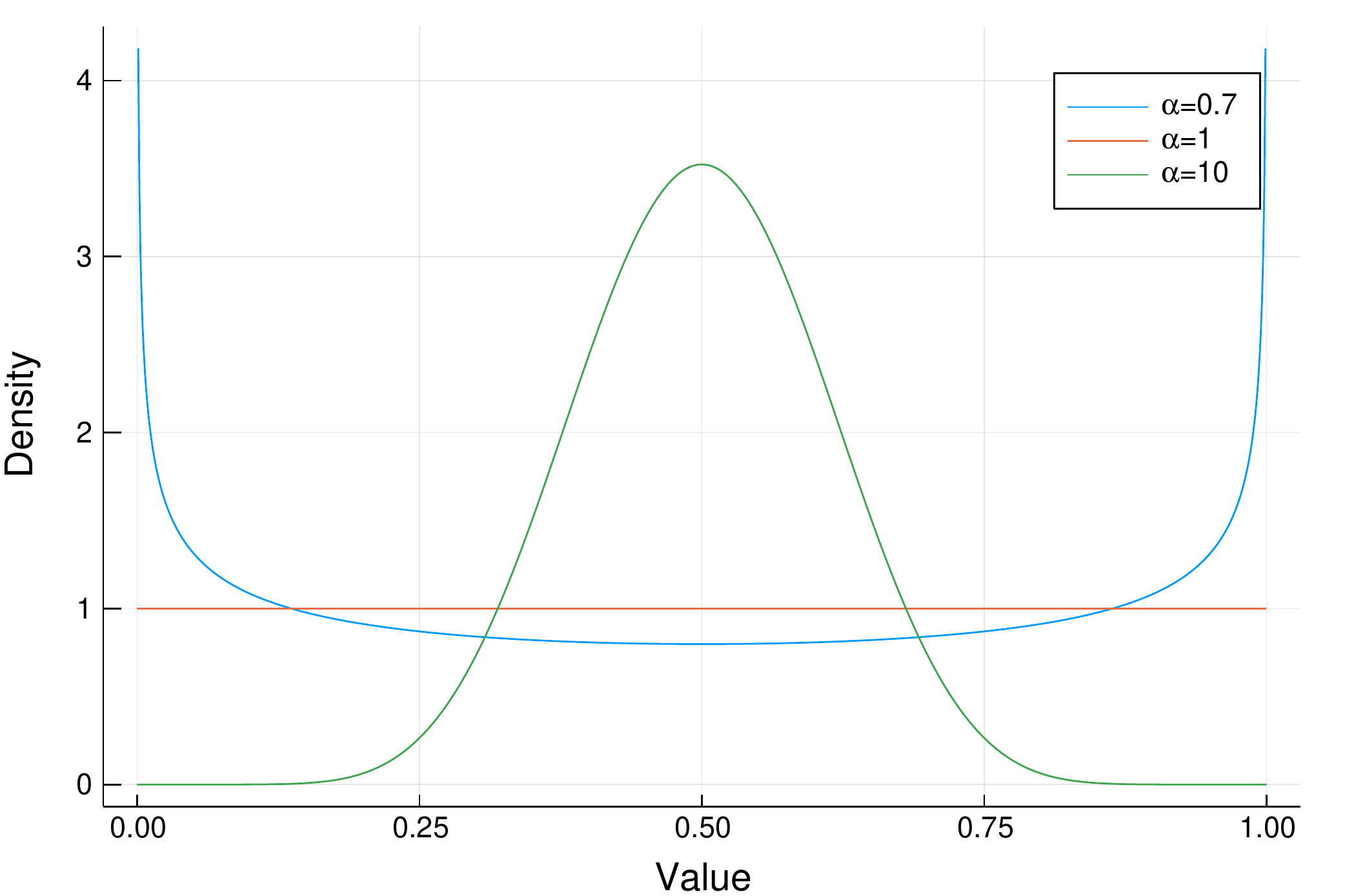} 
\label{beta} 
\end{figure}

The choice of the hyperparameters for the Dirichlet distribution is important, but the economic model provides some guidance. For example,  the prior for $\bm \Pi$ relates to assumptions about the underlying  distributions that determine the preference for news sources,  $F_u$ and $F_e$. Consider as an example  the  case of two categories and a single group when the Dirichlet distribution for $\bm \pi_{1,:}$  is also a Beta$(\alpha,\alpha)$ distribution.  Figure \ref{beta} plots three cases:  $\alpha>1,\alpha<1,$ and $\alpha=1$.  Values of $\alpha < 1$ are associated with values of $\bm \pi_{1,:}$ that are on the corners of the simplex. This implies that the researcher believes that observed heterogeneity is tightly linked with unobserved heterogeneity, or that  members of a group are likely to choose the same source of information.  Values of $\alpha >1$ are associated with  values of $\bm \pi$ in the center of the simplex, which implies that group membership is not highly correlated with unobserved heterogeneity. The case $\alpha=1$ is an uninformative prior, with equal weight everywhere in the simplex. In the economic model, the unobserved heterogeneity is due to information choice. Hence specifying $\alpha<1$ means that the researcher believes that  the group specific cost of acquiring information is more variable than that of the individual specific benefit, so individuals in a certain group are likely to choose the same information. A similar discussion relates the prior for $\bm \beta$.  Specifying a prior with $\eta<1$ would reflect the econometrician's prior that all individuals who choose the same source of information are likely to respond the same way to each question. Although it is constant in this example, the hyperparameter vector need not be constant. The econometrician can specify  higher probabilities on certain areas of the simplex of all possible beliefs over question responses.

  In the examples that follow, we impose $\alpha_{g,k} = 1$ for all $g$ and $k$, which is an uninformative prior for the relationship between group membership and information acquisition. We impose that $\eta_{jkm}=1$ for $k\neq m$, and $\eta_{jkk}= 10$ otherwise. Effectively, we assume that each information source is associated with a correctness score on one response that is higher than any other information source  for at least one question in the survey. For example, for the Michigan Survey of Consumers, we assume that the first information source induces a higher probability on the first categorical response compared to other responses for all questions. The first question asks  respondents about their financial condition compared to a year ago and the first categorical response to this question is `better now'. We thus interpret  the first belief type estimated in multiple MCMC procedures  to be the `optimistic' one. The dynamic model has $K$ more parameters to estimate, $\sigma^2_k$, which has its own priors. Otherwise it is the same; we impose the same kind of prior restrictions on $\bm\beta$. 
  
An additional issue relating to identification is that the frequency response matrix cannot have columns or rows that are near zero. This effectively rules out including questions with very rare responses; in some applications, responses are dropped or combined with more common responses. This condition, however, can be checked ahead of estimation.

\subsection{The Choice of $K$}
 Many solutions have been proposed to determine the number of latent classes in continuous data, but each has some shortcomings. Calculating the Bayes Factor for this type of hierarchical mixture model is computationally challenging due to the intractable integral over the parameter space required to compute the integrated likelihood. Other methods that integrate model selection into the MCMC procedure are complex and require specification of pseudo priors. Cross-validation is popular in the machine learning literature, but it is sensitive to the number of folds and the splits taken. Analyses on choosing the number of latent classes in discrete data are more limited. Early work by \citet{mchugh:56} and \citet{goodman:74} provided insightful results for simple latent class models without a hierarchical structure. Simulation studies for hierarchical latent class models, such as those in \citet{airoldi2006discovering},  have considered different criteria with no clear recommendation. 

 We use insights from both matrix factorization and statistical model selection to guide the choice of $K$, the number of  belief types induced by differing sources of information. We first check that there are enough `degrees of freedom' for a non-negative matrix factorization of $\bm Y_S$. This is a simple counting exercise: the number of independent observations in $Y$ is $G \times (L -J)$; the number of free parameters for $\bm H_S$ and $\bm W_S$ is, respectively, $K \times (L-J)$, and $G \times (K-1)$. This bound is based on counting entries in the frequency matrix alone. Additional information can loosen the bound. For example, \citet{anandkumar2012spectral}  derived an algorithm that also uses  third moments of the data to estimate LDA. To rule out under-identification would require  $G(L-J)\ge K(L-J)+G(K-1)$, or
\[ K \leq G- \frac{G(G-1)}{L+G-J} \] 
If $G=2$, $J=2$ and $L=4$, then the condition requires $K=1$ which is no longer a mixture model.  However, if $G=5$, $J=4$ and $L=20$, the condition $K\le 5-20/21$ tolerates $K$ up to 4 classes. Though we are ultimately interested in parametric estimation of LDA-S, this algebraic result serves as a useful benchmark. 

We run Monte-Carlo simulations to examine the implications of this identification condition in practice. Data are simulated using LDA-S as the DGP under two settings, one where the counting rule is satisfied and the others where it is violated.  We evaluate the posterior mean of $\bm \beta^1_{1,:}$  from the MCMC procedure. In the setting where the counting rule is satisfied, we find that the correlation between the posterior mean of $\bm \beta^1_{1,:}$  and the true  $\bm \beta^1_{1,:}$, averaged across 500 replications, approaches 1 as the sample size increases. This suggests that LDA-S is recovering the unique solution to the problem.   On the other hand, when the model is underidentified and our counting rule fails, the model does not yield  estimates that improve as the sample size increases because  there are multiple solutions by design. This suggests that the counting rule is a useful check of identifiability.  

In analysis of continuous data, it is well known from the Eckart-Young theorem that the best rank $K$ approximation of $\bm Y_S$ is spanned by the eigenvectors corresponding to the $K$ largest eigenvalues of $\bm Y_S\bm Y_S^T$. A similar idea can be expected to hold when $\bm Y_S$ is a function of discrete data.  The eigenvalues of $\bm Y_S\bm Y_S^T$ can be used to generate a scree plot. If $\bm Y$ has rank $K$, there should be $K$ eigenvalues that are significantly larger than the rest. This  can be used to guide the number of classes needed to explain a prescribed fraction of the variance of $\bm Y_S$. Though this method is informal, it is useful in applications because this can be checked ahead of estimation. 

 In the applications below, we use the BIC to approximate the Bayes factor. Let $L(\hat{\theta}_k)$ be the maximum likelihood value of the data and $\theta$ be the set of parameters in the model. For a model with $k$ classes, $ BIC_k =  \left[-L(\hat{\theta}_k) + \frac{1}{2} p_k \log(N) \right] $ 
where $\hat{\theta}_k$ is the maximum likelihood value of the parameters and $p_k$ is the number of model parameters when the number of classes in the model is $k$. We approximate the maximum likelihood value of the parameters using the posterior mean $\tilde{\theta}_k$ from the MCMC draws. The BIC considered is
\[ \tilde{BIC}_k =  \left[-L(\tilde{\theta}_k) + \frac{1}{2} p_k \log(N) \right] \] 
where the sample size $N$ is the number of individuals. In simulations, this approximated BIC criterion reliably selects $K$ for models where the order condition for identification is met, even as we vary the form of the priors. Furthermore, the $K$ chosen by the BIC tends to coincide with the number of eigenvalues that explain over 90\% of variation in $\bm Y_S$.


\section{Applications}
This section consists of two parts. The first application estimates multiple types of respondents from the Michigan data using an LDA-S model. The goal is to show that while the proportion of optimistic respondents over time matches up well with the published index, the other predominant belief types in the data convey additional useful information on recession recovery and economic uncertainty. The second example uses the data analyzed in \citet{Card1995} to illustrate how the estimated individual-level latent class belief types are linked to heterogeneous returns to education. 

 \subsection{Heterogeneous Beliefs in Repeated Cross-Sections: The Michigan Data} 

The Michigan Index of Survey of Consumers (ICS) is a heavily watched indicator of consumer confidence. The index is constructed from monthly survey responses of approximately 500 telephoned respondents in continental U.S. Each month, an independent cross-section sample of households is drawn, and some are reinterviewed six months later. As explained on the official website of the survey,
the ICS is constructed from five questions of the survey on the respondent's opinion about current and future economic conditions as follows:
(i) For $j=1,\ldots 5,$ compute the relative scores $X_j$ which is the percentage of respondents giving favorable replies to question $j$, minus the percent giving unfavorable replies, plus 100;
and (ii) After rounding each relative score to the nearest whole number, compute 
ICS =$\frac{1} {6.7558} \sum_{j=1}^5 X_j +2.0 $
where 6.7558 is the value of the ICS in 1966 (the base period). The 2.0 is added to correct for sample design changes in the 1950s.

The ICS is an aggregate index and is often found to have little independent predictive power for real consumer spending. We will argue that there is unexploited information in the survey beyond the ICS, which requires applying a methodology that captures heterogeneity in responses. To make this point, we analyze 204,944 survey responses collected between January 1978 and May 2019. We do not model the repeated interviewing that occurs in the Michigan data. We use LDA-DS to analyze an additional 9 questions in the survey related to sentiment  that are not incorporated in the ICS but have been asked since 1978. A challenge in using this data is missing values, as many survey respondents may not answer all questions. An appeal of LDA-DS is that it does not require each question to have similar response categories, and it does not require dropping or imputing responses that are incomplete, as long as they are not too rare in the sample. 

\begin{figure}
  \caption{LDA-DS Indices for Michigan Consumer Survey Data  \label{michpi}} 

 \begin{subfigure} [] {0.99\textwidth}
  \includegraphics[width=\textwidth]{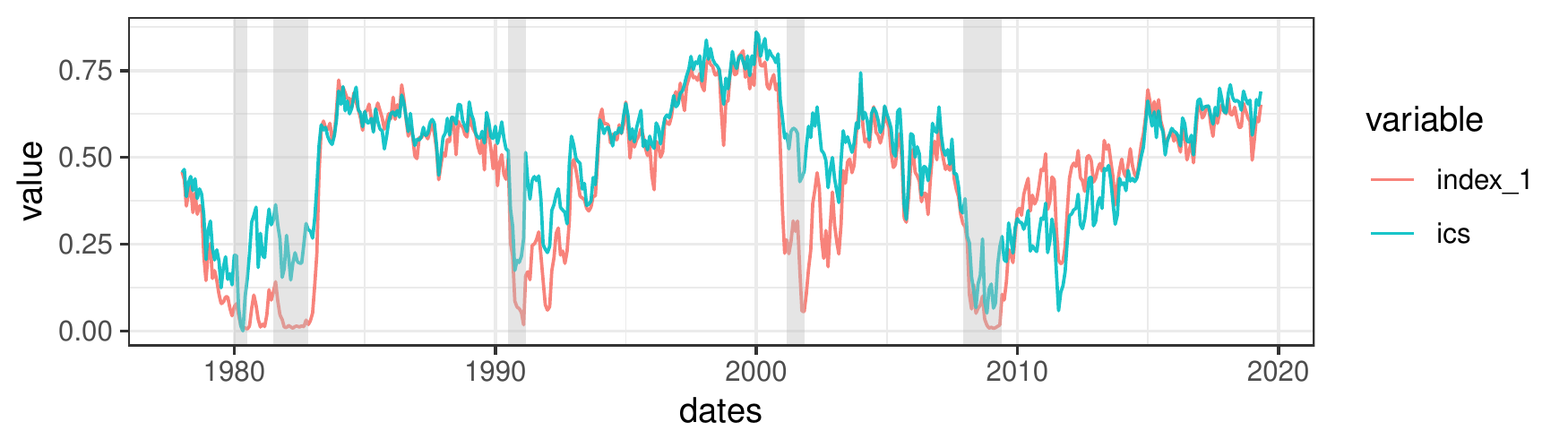}
  \caption{Sentiment corresponding to optimism} 
  \end{subfigure} 
   \begin{subfigure} [] {0.99\textwidth}
  \includegraphics[width=\textwidth]{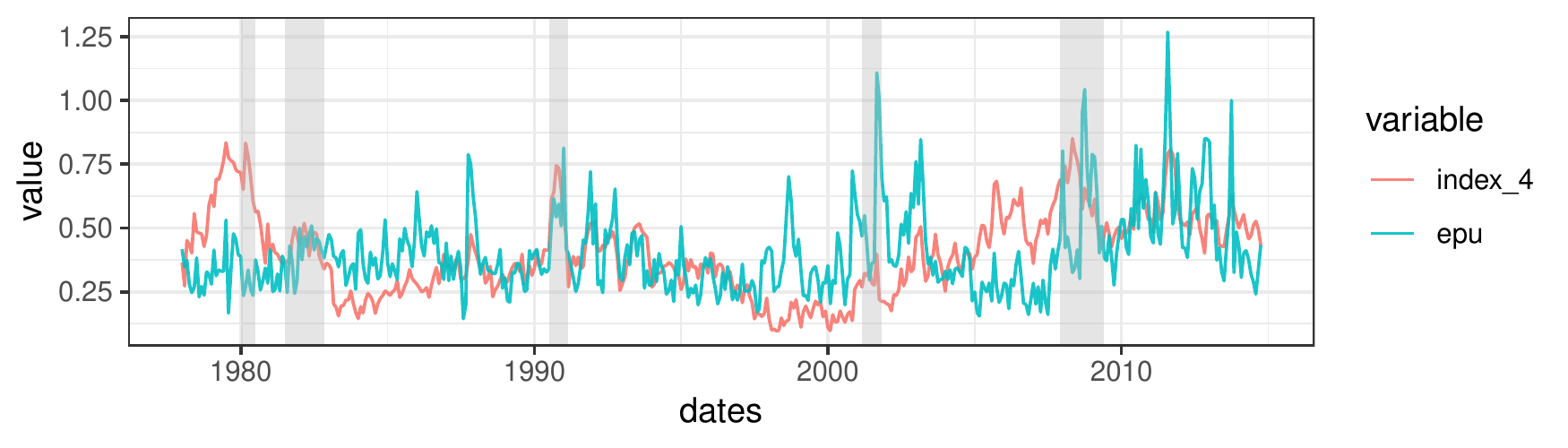}
  \caption{Sentiment corresponding to pessimism} 
  \end{subfigure}
    \begin{subfigure} [] {0.99\textwidth}
  \includegraphics[width=\textwidth]{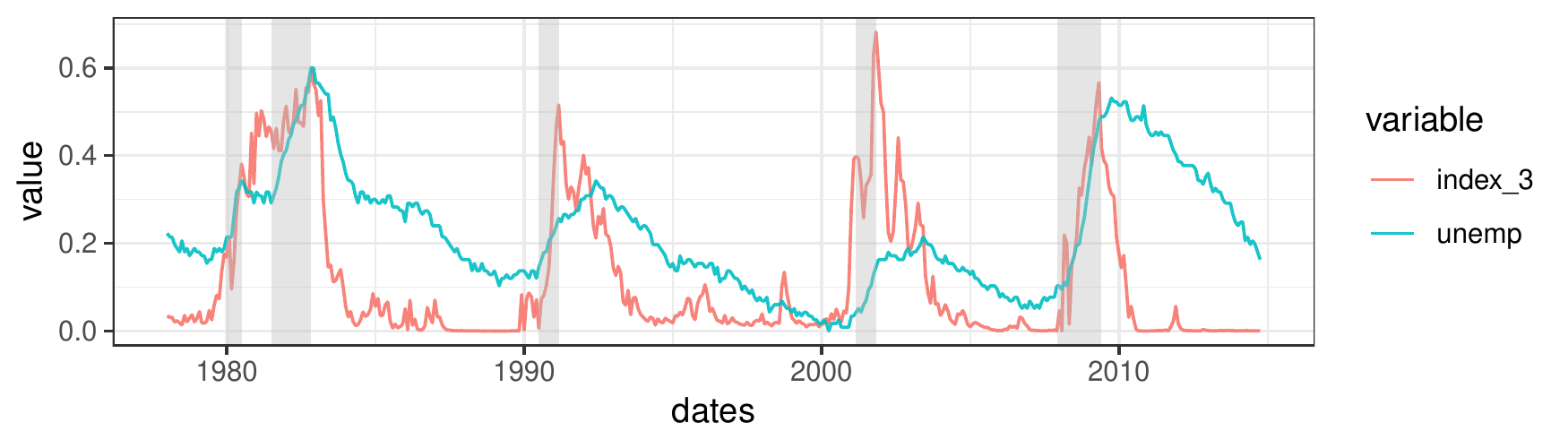}
  \caption{Sentiment corresponding to recession recovery} 
  \end{subfigure}
\end{figure}

\begin{figure} 
\centering
  \caption{Sentiment Types in Michigan Data,  $\bm{\beta}^j_{k,:}$ \label{michbeta}}
 \begin{subfigure} [] {0.42\textwidth}
  \includegraphics[width=\textwidth]{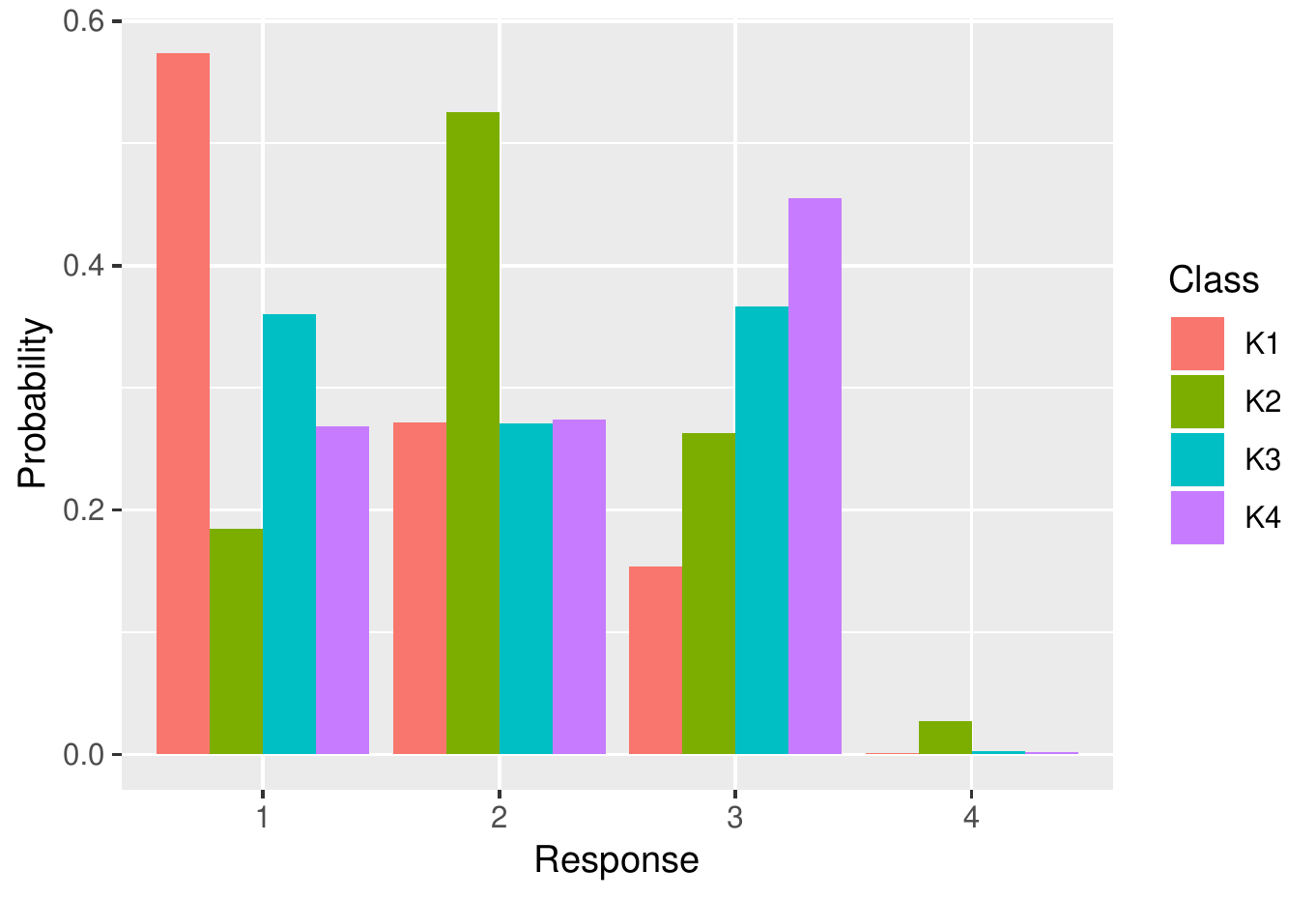}
  \caption{Better (1) or worse off (3) financially than a year ago?} 
  \end{subfigure} 
   \begin{subfigure} [] {0.42\textwidth}
  \includegraphics[width=\textwidth]{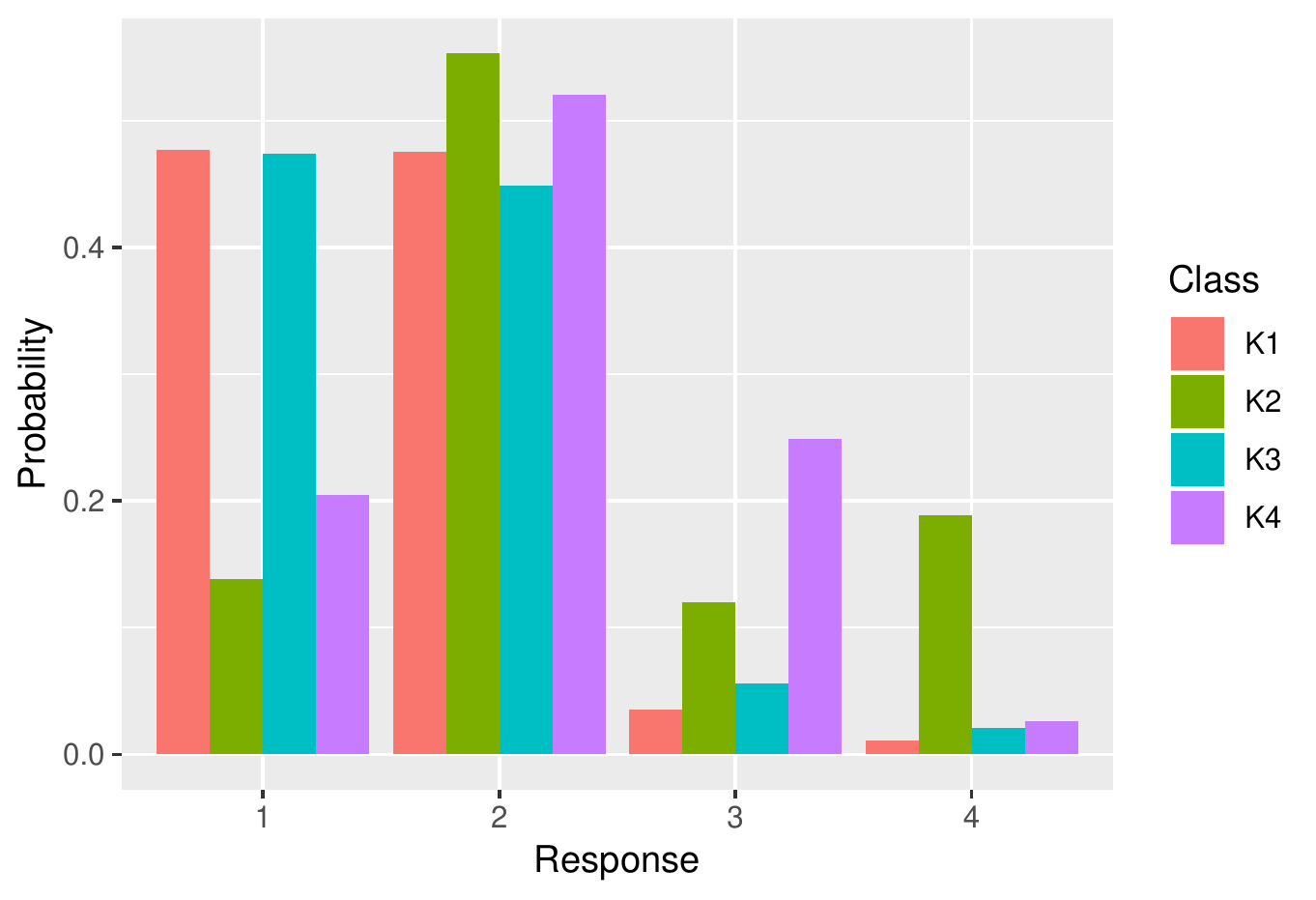}
  \caption{Will you be better (1) or worse (3) financially a year from now? } 
  \end{subfigure} 
   \begin{subfigure} [] {0.42\textwidth}
  \includegraphics[width=\textwidth]{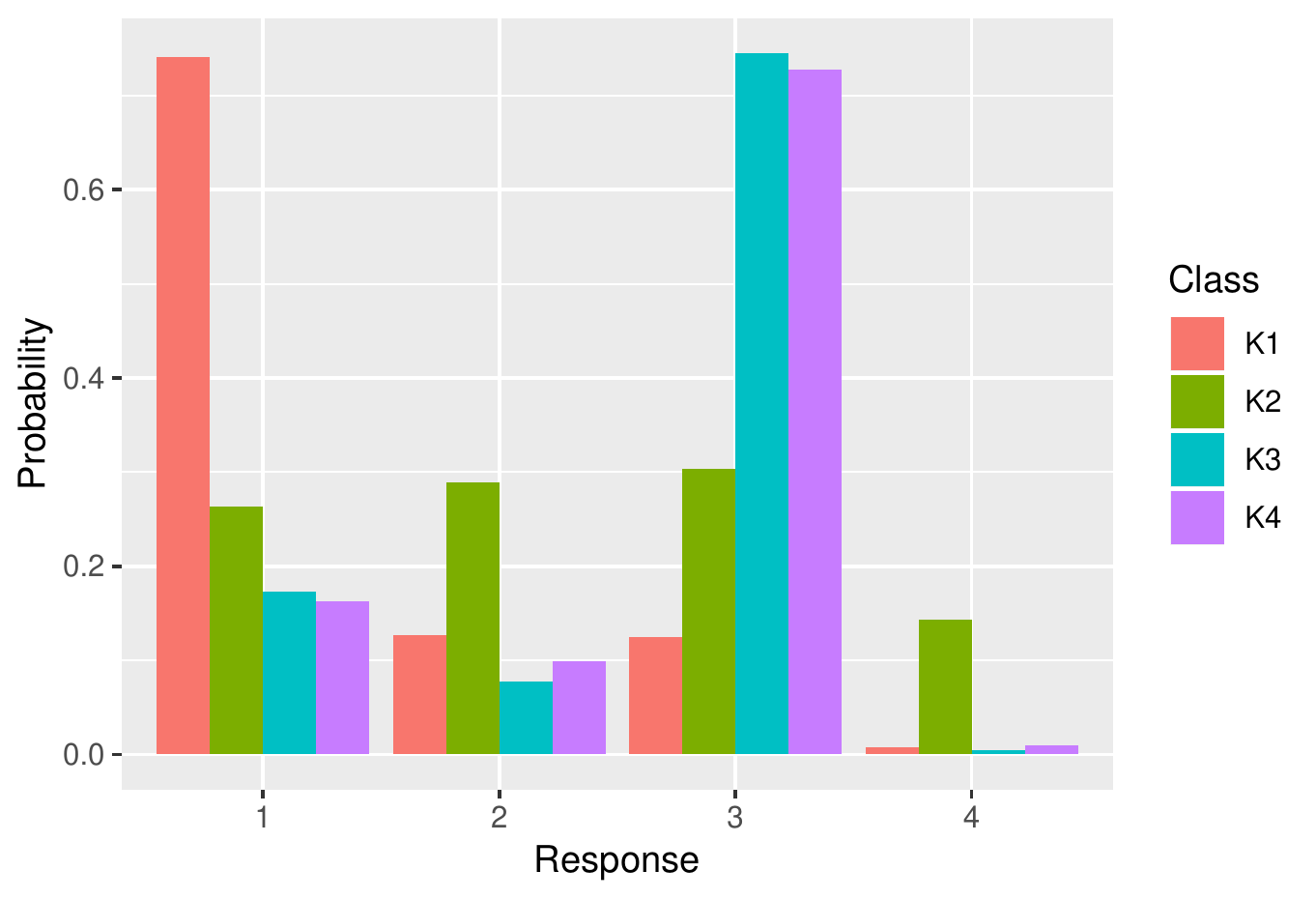}
  \caption{Better (1) or worse (3) business conditions than a year ago? } 
  \end{subfigure} 
    \begin{subfigure} [] {0.42\textwidth}
  \includegraphics[width=\textwidth]{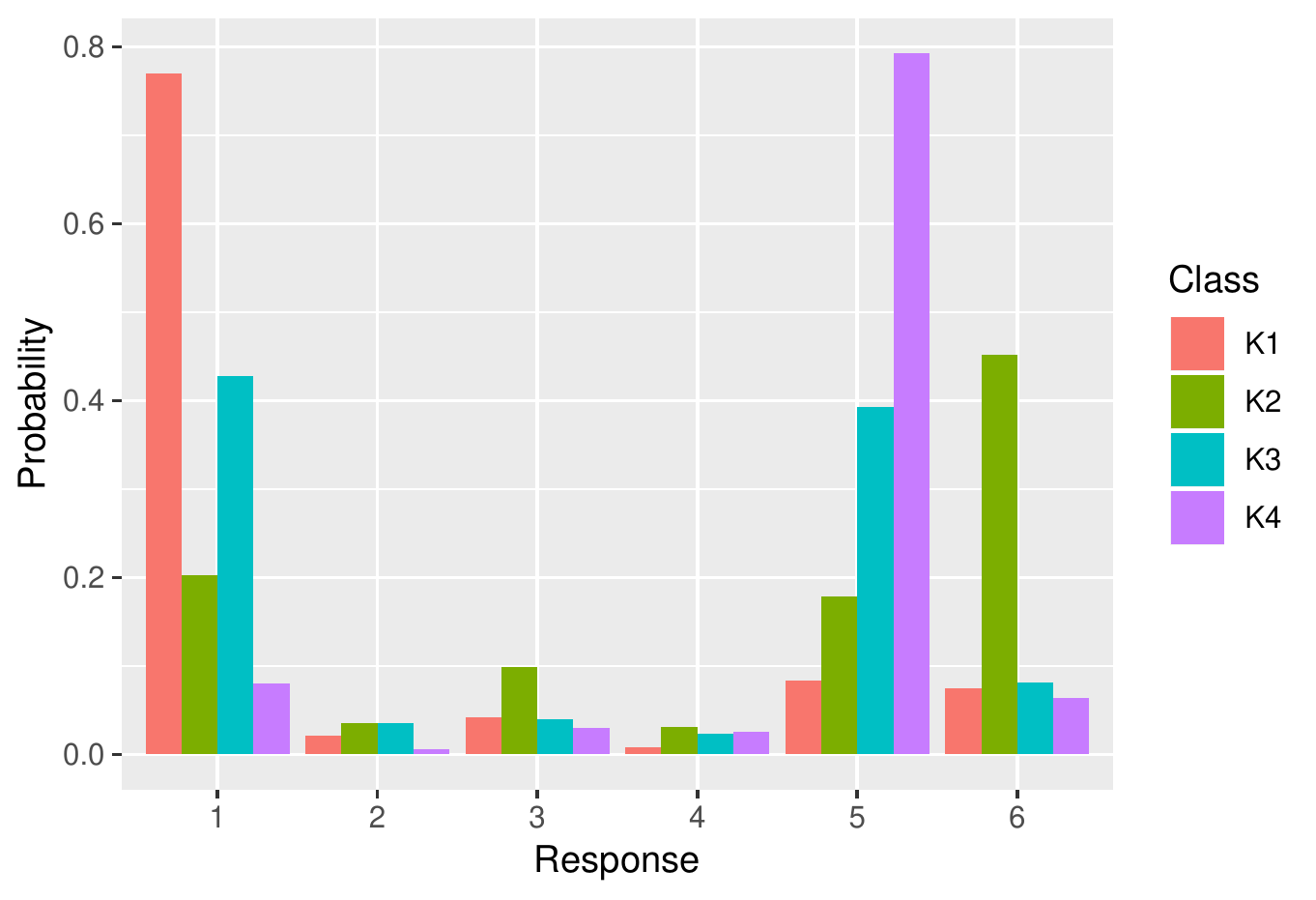}
    \caption{Business conditions good (1) or bad (5) in next 12 months? } 
  \end{subfigure} 
    \begin{subfigure} [] {0.42\textwidth}
  \includegraphics[width=\textwidth]{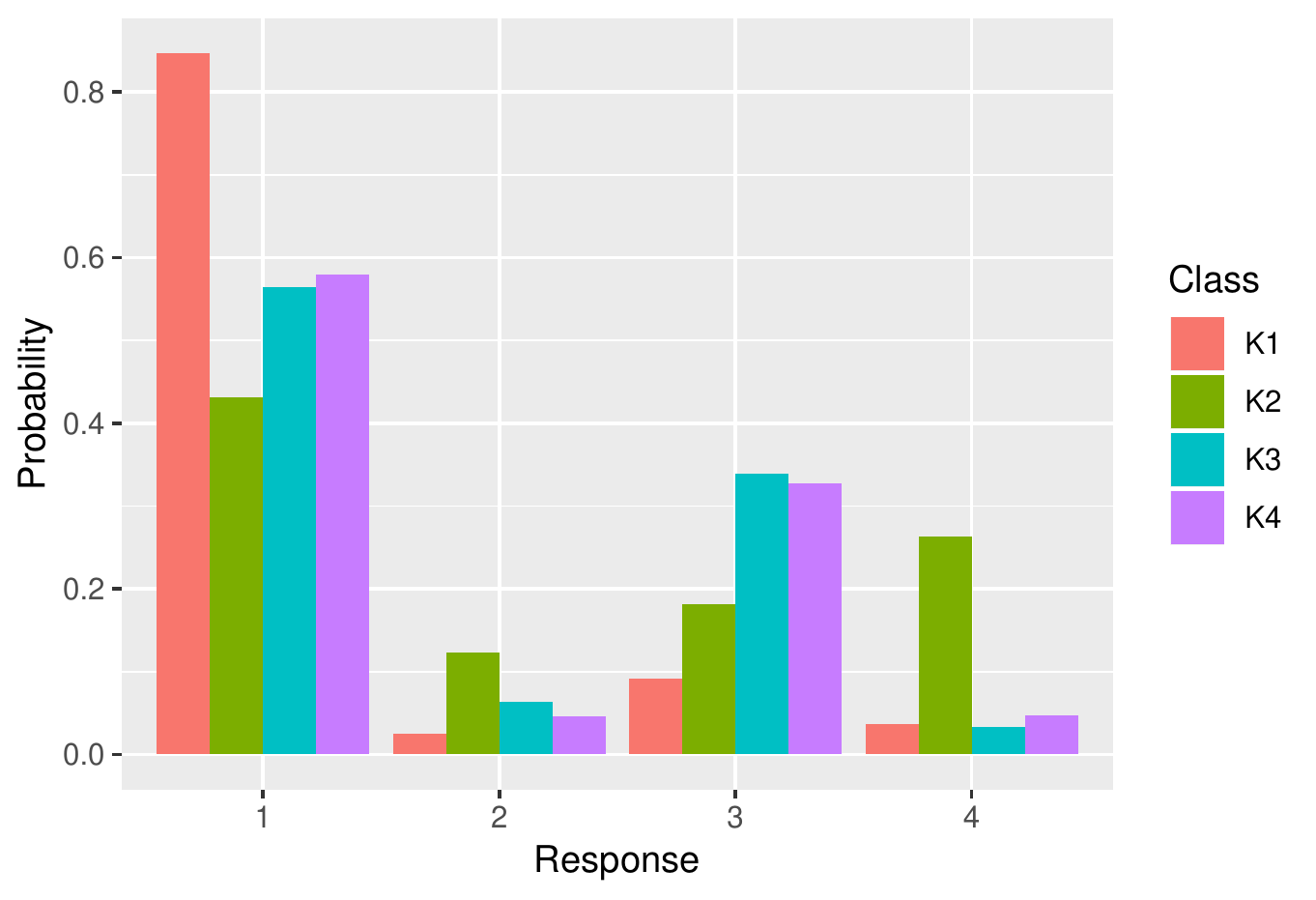}
  \caption{Good (1) or bad time (3) to buy household items?} 
  \end{subfigure} 
  \begin{subfigure} [] {0.42\textwidth}
  \includegraphics[width=\textwidth]{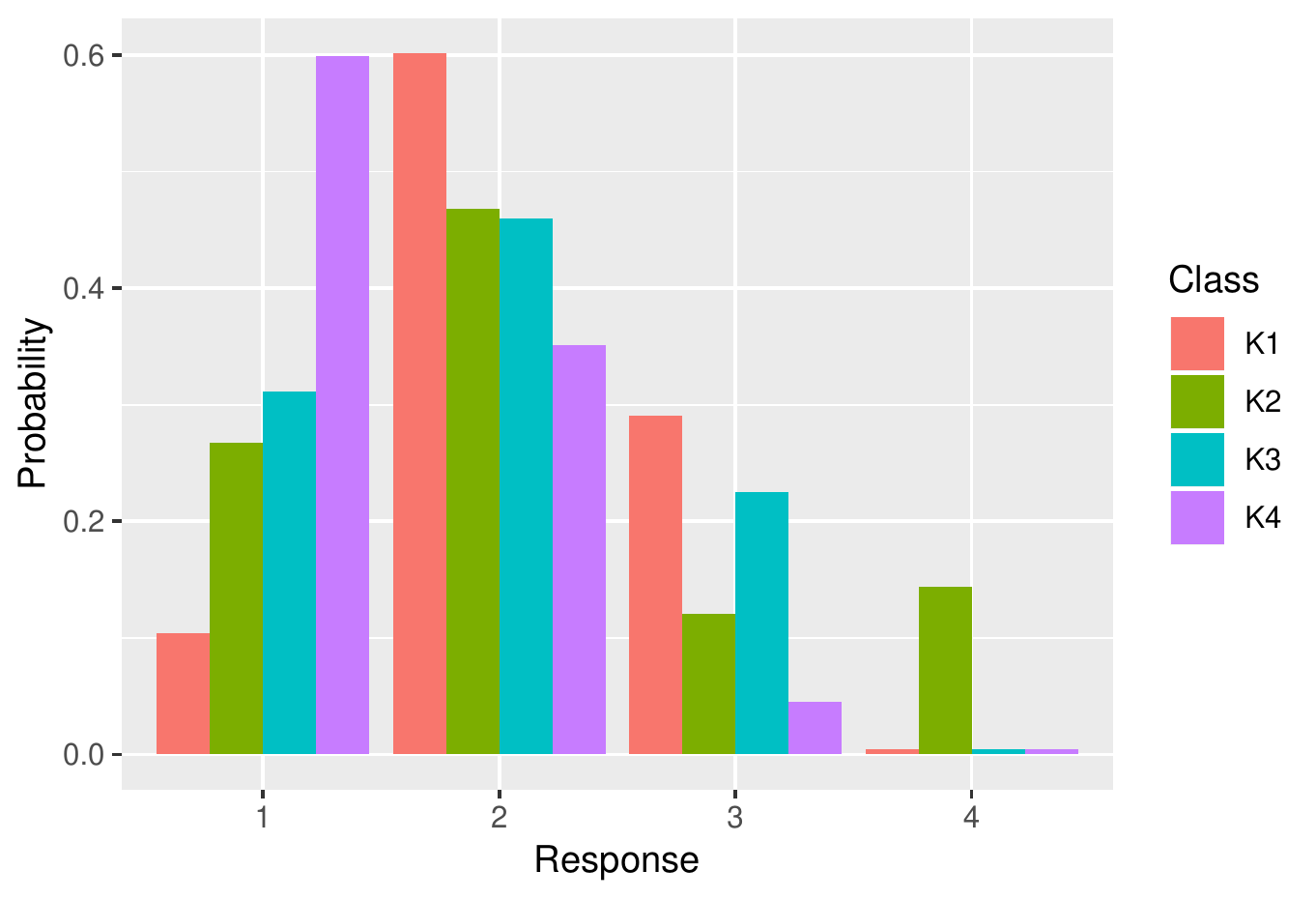}
  \caption{Expect more (1) or less (3) unemployment in the next year? }  
  \end{subfigure} 
\end{figure}

We estimate an LDA-DS model with $K=4$ types. Type 2 captures individuals who are uninformed and are likely to fill out the survey with lots of incomplete or missing responses. This decreases over time presumably due to  changes in survey collection practices. We plot the type proportions $\pi_{1t}$ in Figure \ref{michpi} where $\pi_{1t}$, is the probability that an individual responding at time $t$ is assigned to type 1. A high $\pi_{1t}$ corresponds to a high probability of responding optimistically to the survey questions on sentiment at time $t$. As seen from Figure \ref{michpi}, this index is highly correlated with the rescaled Michigan ICS. Figure \ref{michpi} also plots $\pi_{3t}$ and $\pi_{4t}$  with a rescaled news based uncertainty index and the unemployment rate. While $\pi_{4t}$ often peaks before or at the beginning of the recession, $\pi_{3t}$ peaks towards the end or after the recession. Examining the type-specific multinomial distributions, a few of which are plotted in Figure \ref{michbeta}, suggests that Type 4 is associated with individuals who consume pessimistic news sources, and have a negative response to most questions, reflecting consumers' sentiment that the economy is in bad shape and will continue to be so. This interpretation is supported by the correlation of $\pi_{4t}$ with the news-based uncertainty index, which tracks the proportion of news that contain terms related to uncertainty. The probability of assignment to this `pessimism' profile decreases when times are good in the economy and negative news is less available. In contrast, Type 3 can be understood as a recovery profile since the respondents believe they will likely be better off a year from now, in spite of a reasonably high probability that they have experienced bad times recently. This type is associated with turning points in unemployment.  Estimating more complex forms of heterogeneity beyond an aggregate index is important for identifying and interpreting predominant types of beliefs among respondents to the Michigan survey and how these types relate to aggregate economic conditions. 

 \subsection{Heterogeneous Beliefs and Returns to Education}
 In this subsection, we illustrate how the class assignments $z_i$ can be used to control for unobserved heterogeneity in subsequent regressions. It is well known that unobserved characteristics that are correlated with observed covariates will introduce bias in parameter estimates. With panel data, time-invariant characteristics can be controlled for by differencing or demeaning. These options are not available with cross-sectional data, and instrumental variable estimation is often used. Our approach is to use a high-dimensional set of auxiliary categorical information to control for omitted variable bias and to estimate interpretable heterogeneous effects in a parsimonious way. 

 Our application concerns estimating returns to education, and  education research relies heavily on surveys that solicit subjective assessments.   Using data from the National Longitudinal Survey of Young Men from 1966 to 1976, we consider the regression model 
\begin{equation} y_{i} = \beta_{1i} x_{i} + \beta_2 \bm{w}_i + a_i +e_i\end{equation} 
where $y_i$ is the outcome variable (income), $x_{i}$ is a covariate of interest (education), $\bm{w}_i$ is a small set of important controls, $a_i$ is unobserved heterogeneity, and $v_i= a_i + e_i$ is not observed directly. $\mathbb{E}[e_ix_i] = 0$, but $\mathbb{E}[a_ix_i]\neq0$.   The parameter of interest is the return to education for the men in the sample. In \citet{Card1995},  $\bm{w}_i$ includes covariates such as region, race, ability test scores, and parental education, and proximity to college is then used as an instrument, orthogonal to $a_i$, to estimate returns to education $\beta_{1i}=\beta_1$.

We consider a scenario where a convincing instrumental variable is not available, but we have some auxiliary information about the qualitative beliefs of an individual to control for unobserved heterogeneity in belief type and to estimate heterogeneous effects that are based on an individual's belief type. Our approach uses elements from \citet{Kasahara2009} and \citet{bonhomme-lamadon-manresa}.  However, we not only cluster the data, but also give an economic interpretation to the latent class. We parameterize $a_i$ as a function of individual's belief type:
 \begin{eqnarray*} 
 a_i &=& \sum \limits_{k=1}^K \phi_k {p}(z_i=k) + b_i, \quad E[b_i] = 0.
 \end{eqnarray*}

We further assume that the slope parameters also depend on belief type, so that
 \begin{eqnarray*}
 \beta_{1i} &=& \sum \limits_{k=1}^K \alpha_{k}{p}(z_i=k) +\gamma_i, \quad\quad 
\mathbb{E}[\gamma_i x_i] = 0, \mathbb{E}[\gamma_i]=0.
 \end{eqnarray*} 
Though $p$ is not observed, we can estimate these latent class probabilities if additional data on an individual's beliefs are available. In addition to traditional economic questions on income and education, the U.S. National Longitudinal Survey asks other qualitative questions on beliefs and values. There are in fact hundreds of categorical variables available in the NLSYM survey waves that were not used in \citet{Card1995}. Therefore, in the first step, we estimate a model that allows the probabilities of class membership to vary by groups.  In a second step we use the LDA-S posterior probabilities of membership in each belief type to generate an individual specific intercept, and interact type membership with the slope parameters. In a third step, returns to education are  estimated by least squares regression of 
\begin{equation} 
y_{i} = \sum \limits_{k=1}^K \phi_k {\hat p}(z_i=k) + \sum \limits_{k=1}^K \alpha_{k}\hat{p}(z_i=k) x_{i} + \beta_2 \bm{w}_i +  \epsilon_i 
\end{equation} 
where $\epsilon_i = \gamma_ix_i + b_i + e_i $. If the belief type assignment provides a good proxy for the omitted variable $a_i$, then the bias on $\beta_{1i}$ should be  reduced, and will be eliminated entirely if the belief type captures any relationship that $a_i$ has with $x_i$, or with $y_i$.

 To illustrate, we use 2,830 of the original 3,010 men from  \citet{Card1995} who have Rotter questionnaire data available.   We include (i) 11 variables from the 1976 survey wave that are part of the Rotter questionnaire, a locus of control psychological questionnaire that determined whether or not people believe individual choices determine outcomes or if external factors were to blame. We also include (ii) an additional 3 variables from the 1966 survey, on attitudes about high school and high school courses.  It would be difficult to include all of these variables directly in the regression to measure heterogeneous effects without some method of clustering individuals, since the dimensionality of 14 categorical variables with many different response categories is high. LDA-S overcomes this problem since the information in the categorical variables for an individual is now summarized into interpretable belief type probabilities, which are low in dimension.

We estimate an LDA-S model with $K=3$, where individuals are grouped by the four possible combinations of having a library card interacted with having a single mother at age 14. Table \ref{cardols} contains the results for a replication of \citet{Card1995}'s most basic OLS specification in column (1), as well as the results from estimating Equation 4 in column (2). The mean return to education does not change materially when an individual's belief type is added as a control, which suggests that the estimated belief types are not affecting the bias of the OLS coefficient in this setting. This is similar to what is found in \citet{Card1995}, where the OLS coefficient changes little as additional controls are added beyond the basic specification. 

\begin{figure} [ht] 
\centering 
  \caption{Belief Types in NLSYM Data,  $\bm{\beta}^j_{k,:}$ \label{cardbeta}} 
 \begin{subfigure} [] {0.49\textwidth}
  \includegraphics[width=\textwidth]{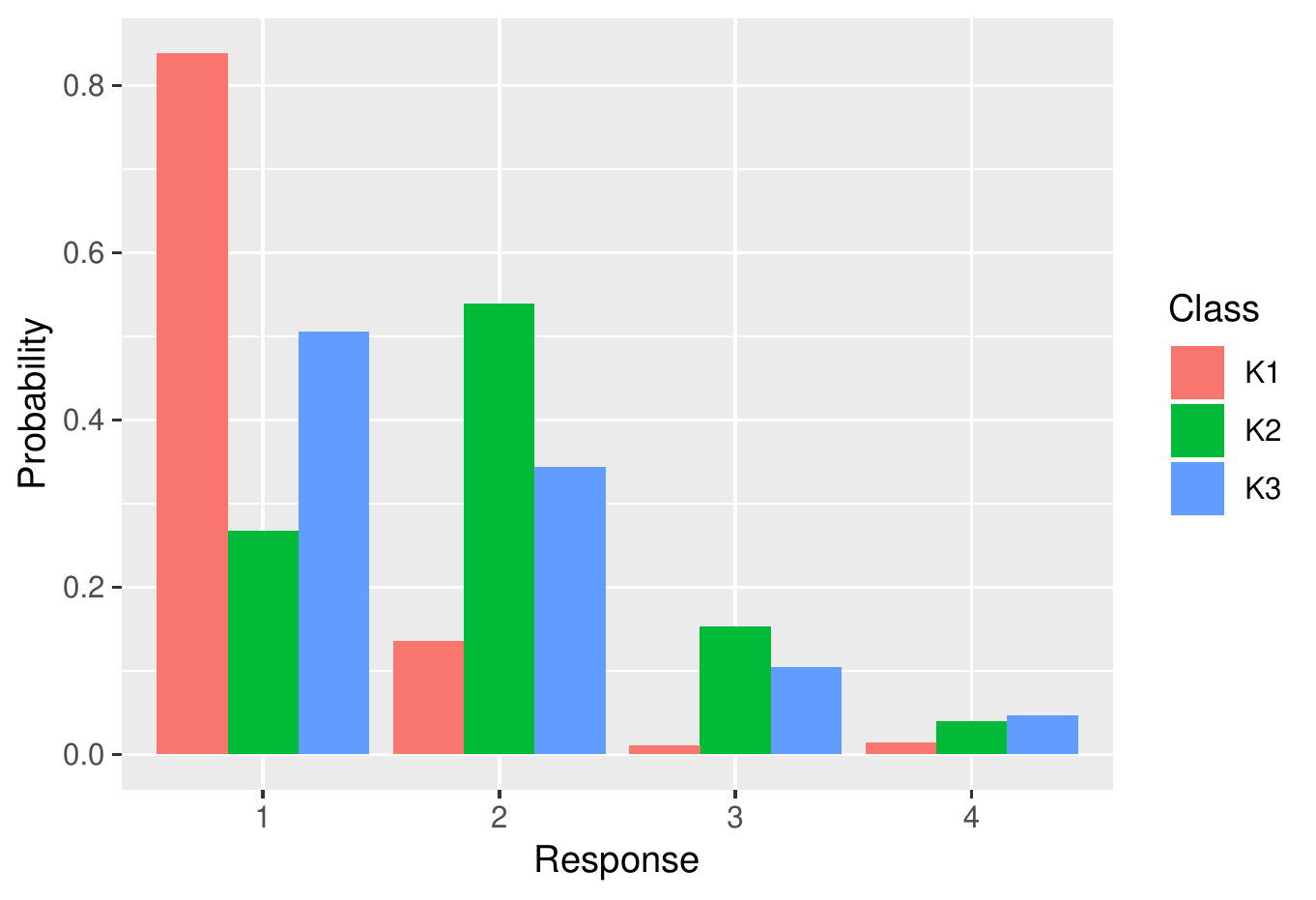}
  \caption{Rotter G: Getting what I want has to do with luck?} 
  \end{subfigure} 
   \begin{subfigure} [] {0.49\textwidth}
  \includegraphics[width=\textwidth]{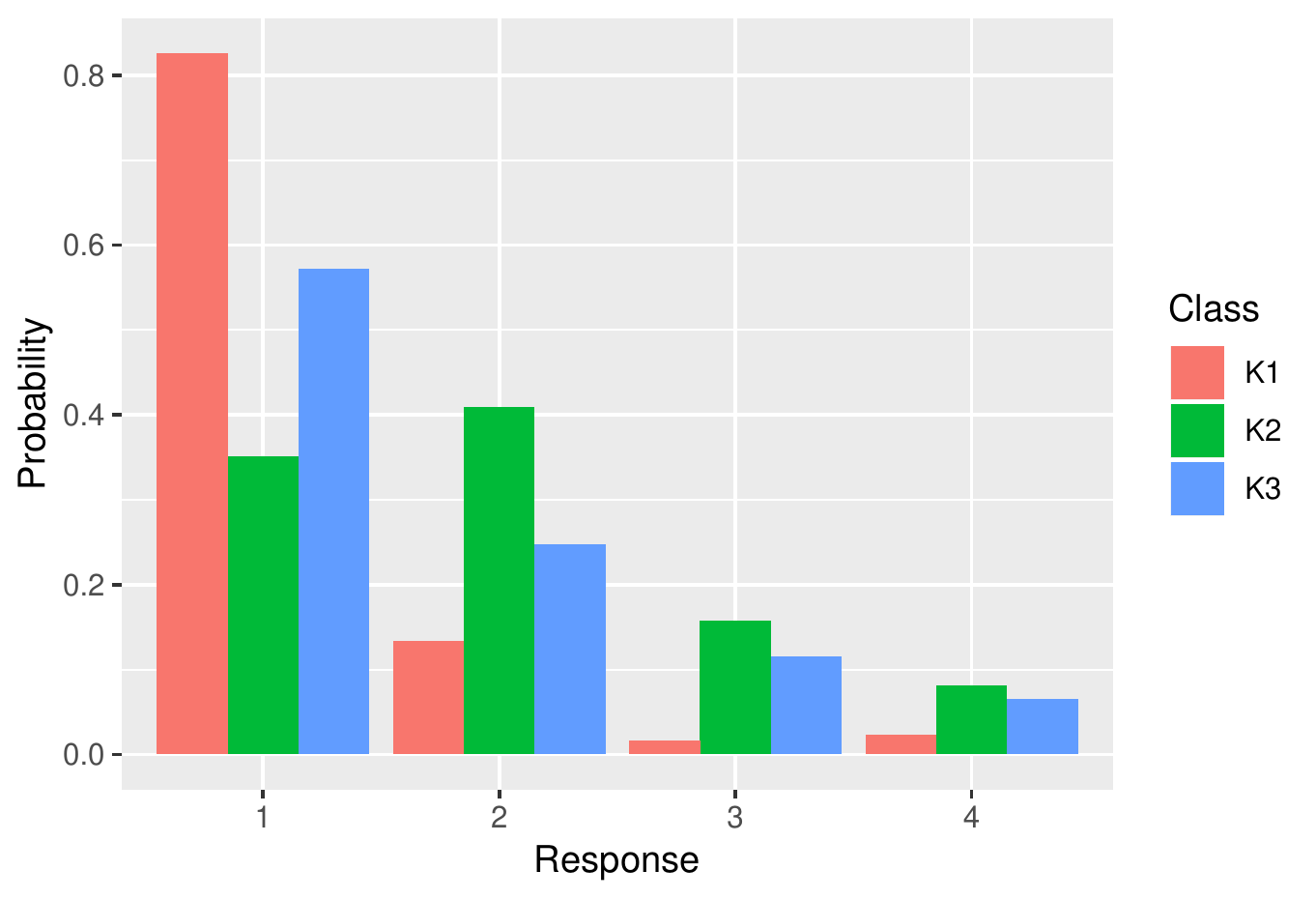}
  \caption{Rotter H: Leadership determined by ability?} 
  \end{subfigure} 
   \begin{subfigure} [] {0.49\textwidth}
  \includegraphics[width=\textwidth]{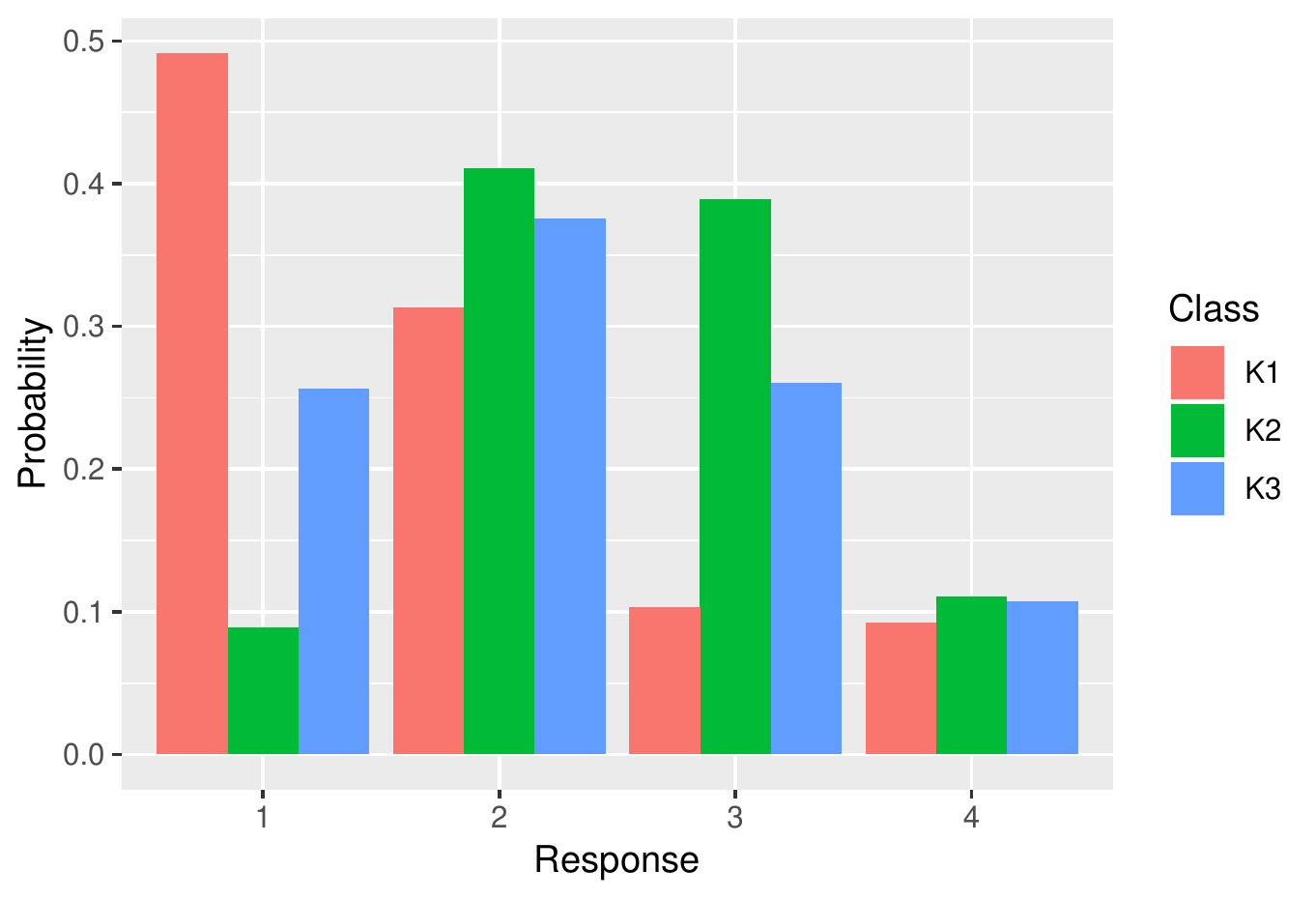}
  \caption{Rotter K: What happens to me is in my control, or not?} 
  \end{subfigure} 
  \end{figure}

 However, there is meaningful heterogeneity in returns to education across belief types. The expected return to education for an individual with  belief type 1 is 5.7\%, while the return of an individual with belief type 2 is 7.9\%. To interpret these differences, we examine $\bm{\beta}^j_{k,:}$. In Figure \ref{cardbeta}, we plot the posterior means of $\bm{\beta}^j_{k,:}$ for the three questions $j$ that vary the most between belief type 1 and belief type 2, as measured by the Rao distance (see \citet{Rao1945}), between $\bm{\beta}^j_{1,:}$ and $\bm{\beta}^j_{2,:}$. Responses 1 and 2 to the survey questions are responses labelled `most internal' and indicate the individual believes success is a matter of internal drive, persistence, and leadership ability. Responses 3 and 4 are labelled `most external', and correspond to individuals believing success is a matter of luck, status, or other factors outside your control. Individuals assigned to type 1,  who have higher mean income but lower returns to education, have absorbed information that has caused them to believe that work ethic and inherent ability leads to success with a higher probability than individuals assigned to type 2. This result is intuitive; individuals who believe that work ethic is important are more likely to make other choices that result in their success with or without a formal education, and so have a lower return to education than individuals who believe external factors determine their success. 

The information that an individual absorbs and incorporates into their beliefs can be an important determinant of their economic choices and outcomes. Examining the returns to education for individuals surveyed in the NLSYM, we do not find evidence that controlling for an individual's belief type reduces bias in the estimate of the mean return to education. However, we do find that belief types correspond to heterogeneous returns to education in a very intuitive way. Analyzing heterogeneity in qualitative beliefs can lead to important insights on the variation in the impact of education on economic outcomes. 
\section{Conclusion} 
This paper proposes a Bayesian hierarchical latent class approach to modeling categorical survey responses. 
We motivate the statistical model using an economic model of information acquisition. We  illustrate using two applications how the estimated belief types can be useful in economic analysis of categorical survey data. 

While traditional face-to-face and telephone surveys have declined,  surveys  can now be easily conducted through the internet and mobile phones. For example, web-based surveys have become an effective way to gauge various types of consumers sentiments during the \textsc{covid19} pandemic. As \citet{callegaro-yang} noted, Survey Monkey alone generates 90 million surveys per month worldwide. One can expect more data of this type to be available. LDA-S can be useful for interpreting and characterizing heterogeneity in such data. Given the close relation of LDA-S to the extensive computer science and statistics literature on mixed membership models, a variety of extensions are possible, including more complex forms of dynamics as in \citet{fox2013mixed}, or integrating latent variable estimates in more complex causal analyses as suggested in \citet{wang2018blessings}. 

For clarity and focus, the paper has used language, motivation, and applications appropriate for categorical survey data on beliefs.   However, there are  many other examples of categorical survey data that are used by economists and social scientists. For example, 
indicators of ownership of assets and housing characteristics from, for example, the Demographic and Health Surveys (DHS) and Living Standard Measurement Surveys (LSMS) are often the only source of data from which  to build social and economic indicators to guide policies. The economic structural model used to derive and interpret LDA-S for qualitative beliefs data can be modified appropriately for other types of economic categorical data.  LDA-S  can be further extended within the mixed-membership framework to allow for  more complex forms of dynamics as in \citet{fox2013mixed}, or integrating latent variable estimates in more complex causal analyses as suggested in \citet{wang2018blessings}.

\newpage

\begin{table}[!htbp] \centering 
 \caption{Returns to Education Estimates} 
 \label{cardols} 
\begin{tabular}{@{\extracolsep{5pt}}lcc} 
\\[-1.8ex]\hline 
\hline \\[-1.8ex] 
 & \multicolumn{2}{c}{\textit{Dependent variable:}} \\ 
\cline{2-3} 
\\[-1.8ex] & LWAGE76 & LWAGE76 \\ 
\\[-1.8ex] & (1) & (2)\\ 
\hline \\[-1.8ex] 
 BLACK & $-$0.187$^{***}$ & $-$0.182$^{***}$ \\ 
 & (0.018) & (0.018) \\ 
 EXP76 & 0.081$^{***}$ & 0.079$^{***}$ \\ 
 & (0.007) & (0.008) \\ 
 EXP762 & $-$0.216$^{***}$ & $-$0.208$^{***}$ \\ 
 & (0.034) & (0.038) \\ 
 SMSA76R & 0.168$^{***}$ & 0.167$^{***}$ \\ 
 & (0.016) & (0.016) \\ 
 REG76R & $-$0.125$^{***}$ & $-$0.127$^{***}$ \\ 
 & (0.016) & (0.016) \\ 
 ED76 & 0.072$^{***}$ & 0.074$^{***}$ \\ 
 & (0.004) & (0.004) \\ 
 Z1 & & 0.24$^{**}$ \\ 
 & & (0.114) \\ 
 Z2 & & $-$0.084 \\ 
 & & (0.106) \\ 
 ED76:Z1 & & $-$0.0165$^{*}$ \\ 
 & & (0.009) \\ 
 ED76:Z2 & & 0.0053 \\ 
 & & (0.008) \\ 
\hline \\[-1.8ex] 
Observations & 2,830 & 2,830 \\ 
R$^{2}$ & 0.282 & 0.284 \\ 
\hline 
\hline \\[-1.8ex] 
\textit{Note:} & \multicolumn{2}{r}{$^{*}$p$<$0.1; $^{**}$p$<$0.05; $^{***}$p$<$0.01} \\ 
\end{tabular} 
\end{table} 


\clearpage

\baselineskip=12.0pt

\bibliography{SurveyBib}

\end{document}